\documentclass[preprint,
superscriptaddress,
pra,
10pt]{revtex4}
\usepackage{graphicx,color}
\usepackage{bm}
\usepackage{hyperref}
\usepackage{amsmath,braket}
\usepackage{amsthm,mathrsfs,amsbsy} 
\usepackage{float}
\usepackage[dvipsnames]{xcolor}

\newcommand{\orcid}[1]{%
  ~\href{https://orcid.org/#1}{\includegraphics[width=8pt]{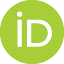}}%
  }

\begin{document}

\title[Strong-Field Ionization of Excited Helium]{Polarization in Strong-Field Ionization of Excited Helium}

\author{A. C. Bray\orcid{0000-0002-8884-4907}}
\address{Department of Physics \& Astronomy, University College London, Gower St., London WC1E 6BT, UK}

\author{A. S. Maxwell\orcid{0000-0002-6503-4661} }
\address{Department of Physics \& Astronomy, University College London, Gower St., London WC1E 6BT, UK}
\address{ICFO -- Institut de Ciencies Fotoniques, The Barcelona Institute of Science and Technology, 08860 Castelldefels (Barcelona), Spain}

\author{Y. Kissin}
\address{Imperial College London - Department of Physics, Prince Consort Rd, Kensington, London SW7 2BW, UK}

\author{M. Ruberti\orcid{0000-0003-0424-3643} }
\address{Imperial College London - Department of Physics, Prince Consort Rd, Kensington, London SW7 2BW, UK}

\author{M. F. Ciappina\orcid{0000-0002-1123-6460} } 
\affiliation{Physics Program, Guangdong Technion -- Israel Institute of Technology, Shantou, Guangdong 515063, China}
\affiliation{Technion -- Israel Institute of Technology, Haifa, 32000, Israel}

\author{V. Averbukh\orcid{0000-0001-7999-0075} }
\address{Imperial College London - Department of Physics, Prince Consort Rd, Kensington, London SW7 2BW, UK}

\author{C. Figueira De Morisson Faria\orcid{0000-0001-8397-4529} }
\address{Department of Physics \& Astronomy, University College London, Gower St., London WC1E 6BT, UK}

\begin{abstract}
We analyze how bound-state excitation, electron exchange and the residual binding potential influence above-threshold ionization (ATI) in helium prepared in an excited $p$ state, oriented parallel and perpendicular to a linearly polarized mid-IR field. Using the ab initio B-spline algebraic diagrammatic construction (ADC), and several one-electron methods with effective potentials, including the Schr\"odinger solver Qprop, modified versions of the strong-field approximation and the Coulomb quantum-orbit strong-field approximation (CQSFA), we find that these specific physical mechanisms leave significant imprints in ATI spectra and photoelectron momentum distributions. Examples are changes of up to two orders of magnitude in the high-energy photoelectron region, and ramp-like structures that can be traced back to Coulomb-distorted trajectories. The present work also shows that electron exchange renders rescattering less effective, causing suppressions in the ATI plateau. 
Due to the long-range potential, the electron continuum dynamics are no longer confined to the polarization axis, in contrast to the predictions of traditional approaches. Thus, one may in principle probe excited-state configurations perpendicular to the driving-field polarization without the need for orthogonally polarized fields.
\end{abstract}
\maketitle

\section{Introduction}

 Above-threshold ionization (ATI) is a strong-field phenomenon in which an atom absorbs more photons than are energetically required for it to ionize. Since its first observation \cite{Agostini}, ATI has been employed to provide unprecedented insight into the interaction of intense
laser radiation with atoms or molecules. The high-energy photoelectrons released into the continuum, together with subfemtosecond resolution, make ATI a powerful tool for probing real-time electron dynamics. Applications of ATI directly associated with imaging and target reconstruction are, for instance, laser-induced electron diffraction (LIED)  \cite{Zuo1996} and ultrafast photoelectron holography \cite{HuismansScience2011}. Such imaging methods combine high photoelectron currents and the possibility of retrieving phase differences without the need for elaborate interferometric schemes. See, e.g., \cite{Faria2020} for a review. 

For that reason, orbit-based models have been used for over two decades due to their huge predictive power. By relating quantum transition amplitudes to electron paths,  which may interfere for a given photoelectron energy, they provide a physical picture of ATI as a laser-induced rescattering process (\cite{Lewenstein1995,Lohr1997,Milosevic1998}; for a review see \cite{Becker2018}).  Attosecond resolution is guaranteed in the rescattering process and ensuing phenomena because it takes place within a fraction of a field cycle, which, typically, is a few femtoseconds.  Established approaches such as the strong-field approximation (SFA)  \cite{Keldysh1965,Faisal1973,Reiss1980} consider the target as a source term and employ a structureless continuum, i.e., field-dressed plane waves. This has allowed the transition amplitudes associated to strong-field phenomena to be written as a laser-dressed Born-type series \cite{Lewenstein1995,Lohr1997,Milosevic1998}, which considerably reduces the numerical effort involved. Besides the easy implementation, an obvious advantage is that the interaction with the core is well-defined. This makes rescattering processes straightforward to identify. Furthermore, the SFA can be easily associated with interfering  electron orbits, if formulated in conjunction with saddle-point methods \cite{Becker2002, Kopold2000, Salieres2001}. 

For that reason, the insight provided by the SFA was vital for identifying the overall structure and interpreting key features in ATI spectra.  The ATI spectrum consists of a low-energy region, up to $2U_p$, where $U_p$ is the ponderomotive energy, and of a high energy
region extending to approximately $10U_p$. The $2U_p$ cutoff corresponds to the `classical' kinetic energy limit for `direct' ATI electrons, which are freed into the continuum via tunnel or multiphoton ionization, and reach the detector without further interaction with the target. The ionization mechanism
that predominates depends on the Keldysh parameter for the system
in question, $\gamma=\sqrt{I_{p}/2U_{p}}$, where $I_p$ is the ionization
potential. A Keldysh parameter of greater than one implies multiphoton
ionization, while tunnelling is dominant if $\gamma\le 1$
\cite{Keldysh1965}.

The high-energy region is mainly dominated by high-order ATI (HATI), which is characterized by a plateau that does not significantly change across
the energy distribution, falling off at the cutoff point of approximately
$10 U_p$. The widely accepted explanation of this feature 
is that the high-energy photoelectrons result from the ionized electron
being driven back by the laser field to scatter elastically off its parent ion.
The rescattered electron is then accelerated further in the field
to energies up to $10 U_p$, which can be derived from
a classical model \cite{Paulus_1994}. Alternatively, the freed electron may recombine with a bound state of the parent ion, which leads to high-order harmonic generation (HHG), or recollide inelastically with the core, releasing other electrons. The latter processes give rise to laser-induced nonsequential double and multiple ionization (NSDI, NSMI) \cite{Faria2011,Becker2012}.
 
Nonetheless, the aforementioned picture is an oversimplification. For example, ATI peaks have a substructure
- termed Freeman resonances - caused by the ponderomotive shifts of
states that produces resonant enhancements \cite{Freeman}. This
effect can also cause broadening of the peaks and even large energy
shifts. Furthermore,  recent discoveries
have cast doubt on the distinction between tunneling and multiphoton regime based on the Keldysh parameter - for instance, photoelectron
spectra obtained from solving the time-dependent Schr\"odinger equation
(TDSE) show features consistent with multiphoton absorption even for
laser intensities that correspond to the tunnelling ionization regime
\cite{Morishita}. In addition to that, specific features of the plateau can vary from system to system. 
Some atomic targets, such as krypton, have a dropping plateau while
others, including xenon, have a flatter plateau. Specific laser intensities
can also cause resonance-like enhancements to the spectra. A variety
of theoretical models have been proposed in recent years to account
for these surprising observations, including Floquet theory \cite{Potvliege},
channel-closing theory \cite{Kopold_2002, Milosevic2007} and models based on the
analysis of TDSE solutions \cite{Muller}. 

Moreover, dynamic effects such as charge migration \cite{Smirnova2009,Mairesse2010,Lepine2014,Calegari2016,Kuleff2016}, bond breaking \cite{Wolter2016}, nuclear degrees of freedom \cite{Karamatskos2019} and multielectron resonances \cite{Shiner2011,Facciala2016} are expected to be important for larger systems.  Extended targets also imply a structured continuum, or a blurred bound-continuum distinction. Modeling such dynamics is not an easy task even for the simplest possible case, i.e., a single-electron continuum under the joint influence of the laser field and a long-range potential. The residual potential blurs the distinction between ``direct" and ``rescattered" as the electron now may be lightly deflected, or undergo a soft scattering by the core \cite{Maxwell2018}. This goes beyond the clear-cut distinction imposed by Born-type approaches such as the SFA \cite{Yan2010,Lai2015}. 
All the above suggests the driving-field polarization as a tool to assess what the SFA leaves out. 
In fact, experimental and theoretical studies have been carried out that demonstrate
the dependence of recollision outcomes - such as HATI or HHG - on
the polarization of the incident radiation \cite{Budil,Dietrich}.
These studies show that the yield is increasingly suppressed as the
ellipticity of the laser pulse is increased. The generally accepted
explanation for this observation is that an elliptical laser will
imbue the ionized electron with a non-zero drift velocity that reduces
the chances of recollision with the parent ion by the electron \cite{Corkum}.
Subsequently, it was demonstrated
that substituting laser polarization for atomic polarization yielded
the same order of magnitude suppression in the production of HHG \cite{Averbukh}.
The present work will extend this line of inquiry to the dependence
of the production of HATI on the atomic polarization. We will focus
for simplicity on the helium atom prepared in an excited state of
$ns^{1}(n+1)p^{1}$ interacting with a linearly polarized field. The
polarization of the \emph{p }orbital can then be oriented along the
laser polarization axis or in the plane perpendicular to it. 
 
In case the atomic polarization is oriented
parallel to the laser polarization, we expect there to be no suppression
to the process outlined above, wherein a continuum electron is driven
by the laser field to scatter off the parent ion. However, if the
atomic polarization is oriented perpendicular to the laser polarization,
ionization in the direction of the laser polarization is precluded
due to the atomic orbital symmetry.
Therefore, ionization is only possible if
the electron acquires a nonzero drift velocity in the plane perpendicular
to the laser polarization. This should greatly reduce the probability
of a successful recollision with the core, as the electron's drift velocity,
unimpeded by interaction with the laser, will displace it away from
the parent ion. Thus, in the high intensity, low frequency regime
for which the recollision model is applicable, we expect the atomic
polarization to significantly affect the HATI spectrum.  On the other hand, residual central potentials may favor ionization or rescattering off the polarization axis, which deviates from the standard predictions. For instance, in \cite{Mauger2010}, rescattering in circularly polarized fields may occur due to the interplay of the laser field and the long-range potential. 
 
In this work, we consider excited helium parallel and perpendicular to the driving-field polarization, for which we calculate photoelectron spectra and momentum distributions. We  use approaches that incorporate the residual binding potential and/or the core dynamics: the B spline algebraic diagrammatic construction (B-spline TD-ADC), the one-electron Schr\"odinger solver Qprop \cite{qprop2,qprop3}, the SFA and the Coulomb quantum-orbit strong-field approximation (CQSFA). They will allow for an assessment of what the simple rescattering picture leaves out. 
The B-Spline TD-ADC is an ab-initio method that solves the 3D atomic many electron time-dependent Schr\"odinger equation for a neutral system \cite{Ruberti2014}. Specifically for the two-electron model studied in this work, it accounts for electron exchange and recollision-induced excitations. In the strong-field context, the B-Spline ADC has been used in high-order harmonic generation \cite{Ruberti2018} and pump-probe spectroscopy \cite{Ruberti2018b}.
The CQSFA is an orbit-based method that incorporates the Coulomb potential and the external laser field on equal footing. It has a high predictive power as it allows for quantum mechanical pathways to be switched on and off at will. In the context of ultrafast photoelectron holography, it led to possibly the most thorough investigations of how holographic patterns form \cite{Maxwell2017,Maxwell2017a,Maxwell2018,Maxwell2018b}.
This work is organized as follows. In Sec.~\ref{sec:theory}, we discuss the theoretical methods used in this work. Subsequently, in Sec.~\ref{sec:results}, we present our results for initial perpendicular and parallel polarized states. Specifically, we investigate angle-integrated spectra (Sec.~\ref{sub:spectra}), and photoelectron momentum distributions (Sec.~\ref{sub:pmds}). Finally, our main conclusions are presented in Sec.~\ref{sec:conclusions}. 

\section{Theory}
 \label{sec:theory}
 
In order to calculate ATI spectra or photoelectron-momentum distributions, we must compute the time-dependent wave function and project it onto an asymptotic continuum state with a well-defined momentum. Below, we will discuss several ways to do so. We will use atomic units throughout, unless otherwise stated, and the dipole approximation. 
 
\subsection{B-Spline algebraic diagrammatic construction}
\label{sec:bspline}

We calculated the ATI spectra of excited, polarised, helium atoms using
the time-dependent (TD) B-spline algebraic diagrammatic construction
(ADC)~\emph{ab initio}~method \cite{Ruberti2014,Simpson2016,Ruberti2018,Ruberti2018b,You2019}.  Within the TD B-spline ADC approach, the 3D many-electron time-dependent
Schr\"odinger equations (TDSE) for the neutral helium atom interacting with the intense mid-IR laser field, given by
\begin{equation}
i\hbar\frac{\partial\left|\Psi_{N} \left( t \right) \right\rangle}{\partial t}\  = \ {\hat{H}}^{N}\left( t \right)\left|\Psi_{N} \left( t \right) \right\rangle,
\label{eq:BSpline2}
\end{equation}
is solved by making the
ansatz 
\[\left|\Psi_{N} \left( t \right) \right\rangle = \alpha_{\text{GS}}\left( t \right)\left|
\Psi_{\text{GS}}^{N} \right\rangle + \sum_{j}^{}{\alpha_{j}\left( t \right)\left| \Psi_{j}^{N} \right\rangle}\label{eq:ADCansatz}
\]
for the time-dependent many-electron state of neutral helium. In 
where \(\left| \Psi_{\text{GS}}^{N} \right\rangle\) represents the
ground state of neutral helium, while the basis states
\(\left|\Psi_{j}^{N} \right\rangle\) refer to the correlated
many-electron configuration states of the ADC theory
for~\emph{N}-electron neutral systems \cite{Ruberti2019}.

The total time-dependent Hamiltonian in Eq.~(\ref{eq:BSpline2}) for the time-evolution of the system interacting with the pulse reads
\begin{equation}
\hat{H^{N}(t)}=\hat{H_{0}^{N}}+\hat{D_{z}}E(t).\label{eq:SEN1}
\end{equation}
Here $\hat{H_{0}^{N}}$ is the field-free many-electron ADC Hamiltonian
describing the neutral system. The laser-atom interaction is described
in the length gauge as $\hat{D_{z}}E(t)$,
where the z component of the dipole operator is $\hat{D_{z}}=\sum_{j=1}^{N}\hat{z_{j}}$
and the summation over the $j$ index runs over all the $N$ electrons of
the atom.

The single-particle basis set used in this approach 
consists of spherical harmonics \(Y_{l,m}\left( \theta,\varphi \right)\)
for the angular part and B-spline
functions~\emph{B\textsubscript{i}}(\emph{r}) for the radial coordinate.
The single particle basis functions used in this calculation are
therefore expressed as
\begin{equation}
\psi_{i,l,m}\  = \frac{B_{i}\left( r \right)}{r}\ Y_{l,m}\left( \theta,\varphi \right).\label{eq:BSpline1}
\end{equation}

The initial excited states used in the present simulations are
\(\ket{\Psi_{N} \left( t = 0 \right) }_{\bot} = \ket{1s2p_{x}}\)
and
\(\ket{\Psi_{N} \left( t = 0 \right)}_{\parallel} = \ket{1s2p_{z}}\)
in the perpendicular and parallel set-ups, respectively. 
In this work, we have used the lowest level of the ADC-hierarchy
compatible with a correct description of the strong-field ionization of
excited helium by the mid-IR laser pulses, i.e. ADC(1). Within ADC(1), the
configuration manifold used to describe the ionisation of helium by the
laser pulses, via solving the TDSE, consists of the singly excited
one-hole--one-particle $(1h-1p)$ configurations. The $1h-1p$ manifold
allows one to describe ionization as well as excitation of helium atom in
any of the singly-excited bound states. The time propagation of the unknown
coefficients~\(\alpha_{j}\left( t \right)\) of the B-spline ADC(1)
many-electron wave-function is performed by means of the
Arnoldi--Lanczos, algorithm \cite{Ruberti2018,Ruberti2018b,Ruberti2019b}. The B Spline ADC within ADC(1) has been successfully used in the strong-field regime to model the intensity-dependent interference minimum that is present in the high-order harmonic spectra of $CO_2$ \cite{Ruberti2018}.
\subsection{One-electron models}

The methods below approximate the system by a one-electron system, whose evolution is given, in atomic units, by the single-electron time-dependent Schr\"odinger equation (TDSE)
\begin{equation}
i\partial_t|\psi(t)\rangle=H(t)|\psi(t)\rangle, \,\label{eq:Schroedinger}
\end{equation}
 where the Hamiltonian
 \begin{equation}
    H(t)= H_a+H_I(t)
    \label{eq:1e-tdse}
 \end{equation}
 describes the joint influence of the binding potential and the external field. Thereby,
 \begin{equation}
H_a=\frac{\hat{\mathbf{p}}^{2}}{2}+V(\hat{\mathbf{r}})
\end{equation}
gives the field-free one-electron atomic Hamiltonian and $\hat{\mathbf{r}}$ and $\hat{\mathbf{p}}$ denote the position and momentum operators, respectively. The coupling with the field is given by the interaction Hamiltonian $H_I(t)=\hat{\mathbf{r}}\cdot \mathbf{E}(t)$ and $H_I(t)=\hat{\mathbf{p}}\cdot \mathbf{A}(t)+\mathbf{A}^2/2$ in the length and velocity gauges, respectively, where $\mathbf{E}(t)=-d\mathbf{A}(t)/dt $ is the electric field of the external laser field and $\mathbf{A}(t)$ the corresponding vector potential.
 
Eq.~(\ref{eq:1e-tdse}) equation is either solved numerically using the freely available software Qprop \cite{qprop2,qprop3} or semi-analytically using the SFA or the CQSFA. Throughout, we employ the effective potential 
\begin{equation}
V(\mathbf{r}(\tau))=-\frac{1+f(r(\tau))}{r(\tau)},
\label{eq:pot}
\end{equation}
where 
\begin{equation}
f(r) = a_1 e^{-a_2 r} + a_3 r e^{-a_4 r}+a_5 e^{-a_6 r}
\label{eq:fpot}
\end{equation}
and $r(\tau)=\sqrt{\mathbf{r}({\tau})\cdot \mathbf{r}({\tau})}$. The coefficients are chosen as $a_1=1.231$ a.u., $a_2=0.662$ a.u., $a_3=-1.325$ a.u., $a_4=-1.236$ a.u., $a_5=-0.231$ a.u. and $a_6=0.480$ a.u. \cite{Maxwell2020}. These parameters were obtained by fitting to a numerical potential computed by self-interaction free density functional theory \cite{Tong_2005, Tong1997}.

For Qprop, the TDSE is solved in the velocity gauge, while for the semi-analytic approaches the length-gauge Hamiltonian is used. This is mainly for practical reasons: the numerical solution of the TDSE converges faster for the velocity gauge, while the length gauge gives better results for ATI in the SFA  \cite{Bauer2005} and CQSFA \cite{Faria2020}. For Qprop we refer to \cite{qprop2,qprop3}, while the semi-analytic models will be summarized below. 
In all one-electron models, for the perpendicular initial $p$ state, the valued real $p_x$ orbital is used via the corresponding coherent superposition of states with $m=\pm1$. 

In Qprop it is possible to compute strong field ionization for the initial states $\ket{\psi_{2p_{\pm1}}}$, where the subscript $\pm1$ refers to the quantum number $m$. The `x' oriented state may be computed by the following superposition,

\begin{equation}
    \ket{\psi_{2p_x}}=\frac{1}{\sqrt{2}}\left(  
        \ket{\psi_{2p_{-1}}}-\ket{\psi_{2p_{+1}}}
    \right)
    \label{cohersuper}
\end{equation}

One can simply take the amplitude components (both real and imaginary) for the initial states $\ket{\psi_{2p_{\pm1}}}$ and compute the coherent superposition above (\ref{cohersuper}) to receive $\ket{\psi_{2p_x}}$ amplitude, which is needed to provide the yield for the photoelectron momentum distributions. 

A convenient starting point for the semi-analytic expressions is the Schr\"odinger equation (\ref{eq:1e-tdse}) in integral form, namely
\begin{equation}
U(t,t_0)=U_a(t,t_0)-i\int^t_{t_0}U(t,t^{\prime})H_I(t^{\prime})U_a(t^{\prime},t_0)dt^{\prime}\,
,\label{eq:Dyson}
\end{equation}
where $U_a(t,t_0)=\exp[iH_a (t-t_0)]$ is the time-evolution operator associated with the field-free Hamiltonian, and the time evolution operator
\begin{equation}
U(t,t_0)=\mathcal{T}\exp \bigg [i \int^t_{t_0}H(t^{\prime})dt^{\prime} \bigg],
\end{equation}
where $\mathcal{T}$ denotes time-ordering, relates to the full Hamiltonian $H(t)$ evolving from an initial time $t_0$ to a final time $t$. Alternatively, one may construct the integral equation as
\begin{equation}
U(t,t_0)=U^{(V)}(t,t_0)-i\int^t_{t_0}U(t,t^{\prime})VU^{(V)}(t^{\prime},t_0)dt^{\prime},
\label{eq:Dyson2}
\end{equation}
 where $U^{(V)}(t,t_0)$ is the Volkov time-evolution operator, associated with the Hamiltonian 
 \begin{equation}
H^{(V)}=\frac{\hat{\mathbf{p}}^{2}}{2}+H_I(t).
\label{eq:Hvolkov}
\end{equation}

We wish to calculate the transition amplitude $\left\langle\psi_{\textbf{p}}(t) |U(t,t_0) |\psi_{0} \right\rangle $ from a bound state $\left\vert \psi _{0}\right\rangle $ to a final continuum state $ |\psi_{\textbf{p}}(t)\rangle$ with momentum $\mathbf{p}$. Using Eq.~(\ref{eq:Dyson}) and the orthogonality between continuum and bound states, it can be written as
\begin{equation}
M(\mathbf{p})=-i \lim_{t\rightarrow \infty} \int_{-\infty }^{t }d
t^{\prime}\left\langle \psi_{\textbf{p}}(t)
|U(t,t^{\prime})H_I(t^{\prime})| \psi _0(t^{\prime})\right\rangle \,
,\label{eq:transitionampl}
\end{equation}
 with $\left | \psi _0(t^{\prime})\right\rangle=\exp[iI_pt']\left\vert \psi _{0}\right\rangle $, where $I_p$ is the ionization potential and we have set $t_0=0$. One should note that, apart from considering a single active electron, no approximation has been made in the time propagation described by Eq.~(\ref{eq:transitionampl}). Some approximations and/or further assumptions will be employed next.

\subsubsection{Strong-field approximation}
\label{sec:sfaexc}

The Strong-Field Approximation (SFA), also known as the Keldysh-Faisal-Reiss (KFR) theory \cite{Keldysh1965,Faisal1973,Reiss1980,Amini2019}, can be obtained by constructing a field-dressed Born series using Eq.~(\ref{eq:Dyson2}), and inserting it into Eq.~(\ref{eq:Dyson}). 

Physically, the SFA relies upon the assumption that all bound states apart from the initial state of the system do not need to be considered. That is, the state of the system at a time $t$ can be
written as
\begin{equation}
\left|\psi(t)\right\rangle =\left|\psi_{0}(t)\right\rangle +\int d\mathbf{p}\, b(\mathbf{p},t)\left|\mathbf{\psi_p}\right\rangle, \label{eq:SFA_wf}
\end{equation} 
where $\left|\mathbf{\psi_p}\right\rangle $ are the continuum states,  
and $\left|\psi_{0}(t)\right\rangle $ is the population in the ground state, often written $\left|\psi_{0}(t)\right\rangle = a(t)\ket{\psi_0}$. If the laser field is not intense enough for the ionization
to reach the saturation level, a widespread approximation is $a(t)=e^{iI_{p}t}$, which means that the ground-state depletion can be neglected.

\subsubsection*{Transition amplitudes}

For ATI, the lowest nonvanishing term of the Born-type series is obtained if the full time evolution operator is replaced by the Volkov time evolution operator $U^{(V)}(t,t^{\prime})$ in Eq.~(\ref{eq:transitionampl}). The subsequent term will include a single interaction in Eq.~(\ref{eq:Dyson2}).
This will lead to two terms that give the amplitude for producing a photoelectron with a particular well-defined momentum,  
\begin{equation}
M(\mathbf{p})=M_{\mathrm{Dir}}(\mathbf{p})+M_{\mathrm{Resc}}(\mathbf{p}).
\label{eq:SFA-all}
\end{equation}
The term $M_{\mathrm{Dir}}(\mathbf{p})$ can be physically interpreted as a direct contribution from electrons that reach the detector with no further interaction with the core, while $M_{\mathrm{Resc}}(\mathbf{p})$ is associated with electrons that are driven back to the parent ion by the laser field and are subsequently elastically rescattered.

The expression for the direct term is
\begin{equation}
M_{\mathrm{Dir}}(\mathbf{p}) =\frac{-i}{(2\pi)^{3/2}}\int_{\infty}^{\infty}dtE(t)e^{iS(\mathbf{p},t)}e^{iI_{p}t}\int d\mathbf{r}e^{-i\left[\mathbf{p}+\mathbf{A}(t)\right]\cdot \mathbf{r}}r\cos\theta\psi_{0}(\mathbf{r})\label{eq:SFA_DIR}
\end{equation}
where $\theta$ is the polar angle, 
$H_I(t)$
is the length-gauge atom field interaction 
and the action reads 
\begin{equation}
S(\mathbf{p},t)=\frac{1}{2}\int_{-\infty}^{t}d\tau[\mathbf{p}+\mathbf{A}(\tau)]^2.
\label{eq:SFAaction}
\end{equation}

The rescattered term is given by
\begin{align}
M_{\mathrm{Resc}}(\mathbf{p})
& =-\int_{-\infty}^{\infty}dt\int_{-\infty}^{t}dt'E(t')e^{iI_{p}t'}e^{-iS(\mathbf{k},t)}e^{iS(\mathbf{k},t')}e^{iS(\mathbf{p},t)}\nonumber \\
& \times\int d\mathbf{k}\frac{1}{(2\pi)^{3}}\int d\mathbf{r'}e^{i(\mathbf{k}-\mathbf{p})\cdot\mathbf{r'}}V(\mathbf{r'})\times\frac{1}{(2\pi)^{3/2}}\int d\mathbf{r}e^{-i(\mathbf{k}+\mathbf{A(t')})\cdot \mathbf{r}}r\cos\theta\psi_{0}(\mathbf{r})\label{eq:SFA2}
\end{align}

The three steps in the rescattering process delineated in Eq.~(\ref{eq:SFA2}) are: (i) ionization at time $t$ to a Volkov state
with canonical momentum $\mathbf{k}$; (ii) propagation in the laser
field from $t'$ to $t$, where $t>t'$; (iii) scattering off the ion
to a final Volkov state with canonical momentum
$\mathbf{p}$ at time $t$, where $V$ can be a model potential for
the ionic system in question. It is very common to calculate the SFA transition amplitudes employing saddle-point methods, as they simplify the numerical effort and provide a very intuitive, clear picture in terms of rescattering electron trajectories (for details see, e.g., \cite{Faria2002}). However, the standard saddle-point method selects the rescattering processes that are exactly along the driving-field polarization. This will be problematic for initial bound states aligned perpendicularly to the field polarization. 
A discussion of how to compute these matrix elements explicitly and how to implement  saddle-point methods in the present context is provided in the appendix. 

For early discussions of how to incorporate rescattering in ATI see Refs.~\cite{Lewenstein1995} and \cite{Lohr1997}. Further detail about the SFA and strong-field ionization is provided in the tutorial \cite{anatomy} and in the review articles \cite{Popruzhenko2014JPB,Becker2018}.  Introducing the term $M_{\mathrm{Resc}}(\mathbf{p})$  is also known as the ``improved strong-field approximation" \cite{Hasovic}. 

\subsubsection*{Bound-state transitions\label{subsec:Bound-state-transitions}}
The standard formulation of the SFA method assumes that the active
electron occupies a specific ground state and that it can only transition
into continuum states through interaction with the laser. The influence
of other bound states on the dynamics of the system is typically not
considered. 

However, if the initial state of the system is excited
to a point where it is within energetic reach of other excited states,
this assumption may become  problematic. In the case of the helium
atom excited into the 2\textsuperscript{1}\emph{P} state, particular
attention should be paid to the transition dipole between the 2\textsuperscript{1}\emph{P}
state and the 2\textsuperscript{1}\emph{S} state. Here, the energy
gap is relatively small, approximately 0.0221 a.u (0.6 eV), and the transition
dipole moment is large, approximately 2.9 a.u. In principle, other possible
bound-state transitions, such as 2\textsuperscript{1}\emph{P} -- 1\textsuperscript{1}\emph{S
}or 2\textsuperscript{1}\emph{P} -- 3\textsuperscript{1}\emph{S }can
be more safely ignored as the energy difference that characterizes
these transitions is significantly larger. Excitation has also been incorporated in the SFA in the context of NSDI \cite{Shaaran2010,Shaaran2010a,Wang2012,Hao2014,Maxwell2015,Maxwell2016}. For coherent superpositions of states in the SFA which, however, are only coupled via the continuum see, e.g., \cite{Faria2010}.

To incorporate this into the SFA model above, we modify the system's
state space vector (see Eq.~\eqref{eq:SFA_wf}) as follows:
\begin{equation}
\left|\Psi(t)\right\rangle =a_{2^{1}S}(t)\left|\psi_{2^{1}S}\right\rangle+a_{2^{1}P}(t)\left|\psi_{2^{1}P}\right \rangle+\int d\mathbf{p}\hspace*{0.1cm}b(\mathbf{p},t)\left|\psi_\mathbf{p}\right\rangle .
\end{equation}

This would modify the first term in Eq.~(\ref{eq:SFA-all}) to
\begin{equation}
M_{\mathrm{Dir}}(\mathbf{p})  =\frac{-i}{(2\pi)^{3/2}}\sum_{j}\int_{\infty}^{\infty}dtE(t)e^{iS(\mathbf{p},t)}a_{j}(t)\int d\mathbf{r}e^{-i\left[\mathbf{p}+\mathbf{A}(t)\right]\cdot \mathbf{r}}r\cos\theta\psi_{j}(\mathbf{r}),
\end{equation}
where the index $j$ runs over the excited states to be included in
the calculation. The same modification is made to the expression for
the term $M_{\mathrm{Resc}}(\mathbf{p})$ in Eq.~\eqref{eq:SFA-all}. The model now describes ionization
from a series of strongly coupled bound states. We determine the time-dependent
behaviour of the bound states in a laser field with the system of
coupled equations:
\begin{equation}
\frac{da_{j}}{dt}=I_{p_{j}}a_{j}(t)-\sum_{i\neq j}\mathbf{E}(t)\cdot \mathbf{d}_{j-i}a_{i}(t),\label{eq:coupled_eqs}
\end{equation}
where $I_{p_{j}}$ is the ionization potential for the excited state
$j$ and $\mathbf{d}_{j-i}$ is the matrix element of the dipole transition
from one excited state to another with the orientation of this element
determined by the spatial polarization of the excited orbitals. If
the number of excited states under consideration is two and the quantity
$2\mathbf{E}_{0}\cdot\mathbf{d}_{j-i}$ (where $\mathbf{E}_{0}$ is the maximum
field strength) is much greater than the bound-state energy difference between
the states $I_{P_{j}}-I_{P_{i}}$, then the system of equations in
(\ref{eq:coupled_eqs}) can be solved perturbatively \cite{Ivanov}.

The solution for our helium system is
\begin{align}
a_{2^{1}S}^{0}(t) & =ie^{-i\tilde{I}_{p}t}\sin\left[\int_{0}^{t}\mathbf{E}(t')\cdot\mathbf{d}_{2^{1}S-2^{1}P}dt'\right]\label{eq:a21s}\\
a_{2^{1}P}^{0}(t) & =e^{-i\tilde{I}_{p}t}\cos\left[\int_{0}^{t}\mathbf{E}(t')\cdot\mathbf{d}_{2^{1}S-2^{1}P}dt'\right],\label{eq:a21p}
\end{align}
where $\tilde{I}_{p}$ is the average energy of the two excited states
and the 0 in the superscript indicates that this solution is to zeroth
order. This means that the energy difference is neglected and the states are
considered as degenerate. The condition for this solution $2\mathbf{E}_{0}\cdot \mathbf{d}_{2^{1}S-2^{1}P}\gg I_{P_{2^{1}P}}-I_{P_{2^{1}S}}$
is not valid in the case of a finite pulse at the edges of the envelope,
but remains applicable provided an alternative condition is met: that
$\omega\gg I_{P_{2^{1}P}}-I_{P_{2^{1}S}}$ where $\omega$ is the
field frequency. It can be readily seen from Eqs.~(\ref{eq:a21s}) and
(\ref{eq:a21p}) that the solution for the perpendicular configuration
is 
\begin{align*}
a_{2^{1}S}^{0}(t) & =0\\
a_{2^{1}P}^{0}(t) & =e^{-i\tilde{I}_{p}t}.
\end{align*}
This is because the 2\textsuperscript{1}\emph{P} -- 2\textsuperscript{1}\emph{S
}transition is not possible due to the symmetry constraints of the
dipole transition in the case of the perpendicular configuration.

\subsubsection{The Coulomb quantum-orbit strong-field approximation}
\label{sec:cqsfa}

Instead of constructing iterative schemes around Eq.~(\ref{eq:Dyson}), one of which is the SFA, one may also employ path-integral methods and construct approaches that incorporate the field and the binding potential on equal footing. One of such approaches is the Coulomb quantum-orbit strong-field approximation (CQSFA) \cite{Lai2015,Maxwell2017}. It is constructed from the transition amplitude
\begin{align}
&M(\mathbf{p}_f)=\notag\\&-i \lim_{t\rightarrow \infty}
\int_{-\infty }^{t }\hspace*{-0.2cm}d t'
\int d \mathbf{\tilde{p}}_0 \left\langle  \mathbf{\tilde{p}}_f(t)
|U(t,t') |\mathbf{\tilde{p}}_0\right \rangle 
\left \langle
\mathbf{\tilde{p}}_0 | H_I(t')| \psi
_0(t')\right\rangle, \, 
\label{eq:Mpp}
\end{align}
which is obtained by inserting a closure relation in the initial field-dressed momentum $\mathbf{\tilde{p}}_0=\mathbf{p}_0+\mathbf{A}(t')$ in Eq.~(\ref{eq:Dyson}) and is exact for a one-electron system. Eq.~(\ref{eq:Mpp}) takes the system from the initial photoelectron bound state $\left| \psi_0(t')\right\rangle \,$ to its final momentum state $\ket{\psi_{\mathbf{p}}(t)}=\left|  \mathbf{\tilde{p}}_f(t)\right\rangle \,$, where the tilde indicates field-dressing. The  $\mathbf{\tilde{p}}_0=\mathbf{p}_0+\mathbf{A}(t')$ and $\mathbf{\tilde{p}}_f(t)=\mathbf{p}_f+\mathbf{A}(t)$, where $\mathbf{A}(\tau)$, with $\tau=t,t'$ is the vector potential,  
denote the electron's initial and final field-dressed momenta, respectively. The time-evolution operator $U(t,t')$ is associated to the full Hamiltonian, and thus includes the laser field \textit{and } the binding potential.  We take the interaction Hamiltonian $H_I(t')$ to be in the length, gauge. Applying  time-slicing techniques \cite{Kleinert2009,Milosevic2013JMP} gives
\begin{align}
M(\mathbf{p}_f)&=-i\lim_{t\rightarrow \infty
}\int_{-\infty}^{t}dt' \int d\mathbf{\tilde{p}}_0
\int_{\mathbf{\tilde{p}}_0}^{\mathbf{\tilde{p}}_f(t)} \mathcal {D}'
\mathbf{\tilde{p}}  \int
\frac{\mathcal {D}\mathbf{r}}{(2\pi)^3}\notag\\
&\hspace{2cm}\times  
e^{i S(\mathbf{\tilde{p}},\mathbf{r},t,t')}
\langle
\mathbf{\tilde{p}}_0 | H_I(t')| \psi _0  \rangle \, ,
\label{eq:CQSFApath}
\end{align}
where $\mathcal{D}'\mathbf{p}$ and $\mathcal{D}\mathbf{r}$ are the integration measures for the path integrals \cite{Kleinert2009,Lai2015}, and the prime denotes a restriction. The action and the Hamiltonian in Eq.~(\ref{eq:CQSFApath}) reads as
\begin{equation}\label{eq:stilde}
S(\mathbf{\tilde{p}},\mathbf{r},t,t')=I_pt'-\int^{t}_{t'}[
\dot{\mathbf{p}}(\tau)\cdot \mathbf{r}(\tau)
+H(\mathbf{r}(\tau),\mathbf{p}(\tau),\tau]d\tau,
\end{equation}
and 
\begin{equation}
H(\mathbf{r}(\tau),\mathbf{p}(\tau),\tau)=\frac{1}{2}\left[\mathbf{p}(\tau)+\mathbf{A}(\tau)\right]^2
+V(\mathbf{r}(\tau)),
\label{eq:Hamiltonianpath}
\end{equation}
respectively. In Eqs.~(\ref{eq:stilde}) and (\ref{eq:Hamiltonianpath}), 
the intermediate momentum $\mathbf{p}$ and coordinate $\mathbf{r}$ have been parameterized with regard to the time $\tau$ and  $\mathbf{\tilde{p}}=\mathbf{p}(\tau)+\mathbf{A}(\tau)$. 

One should note that no Born-type expansion was used in the derivation of Eq.~(\ref{eq:CQSFApath}). This implies that the CQSFA cannot be viewed as a field-dressed perturbative series with regard to the binding potential, and that the distinction between direct and rescattered electrons is blurred. 
This is even clearer when Eq.~(\ref{eq:CQSFApath}) is solved using saddle-point methods, i.e., we search for values of $t'$, $\mathbf{r}$ and $\mathbf{p}$ that render the action (\ref{eq:stilde}) stationary. This gives 
\begin{equation}
\frac{\left[\mathbf{p}(t')+\mathbf{A}(t')\right]^2}{2}+V(\mathbf{r}(t'))+\dot{\mathbf{p}}(t')\cdot \mathbf{r}(t')=-I_p,
\label{eq:tunncc}
\end{equation}
\begin{equation}
\mathbf{\dot{p}}=-\nabla_rV(\mathbf{r}(\tau))
\label{eq:p-spe}
\end{equation}
and
\begin{equation}
\mathbf{\dot{r}}= \mathbf{p}+\mathbf{A}(\tau).
\label{eq:r-spe}
\end{equation}
Eq.~(\ref{eq:tunncc}) yields the energy conservation at tunnelling, and Eqs.~(\ref{eq:p-spe}) and  (\ref{eq:r-spe}) give the subsequent electron propagation.

We compute Eq.~(\ref{eq:Mpp}) along a two-pronged contour that starts in at the complex time  $t'=t'_r+it'_i$, goes parallel to the imaginary time axis, i.e.,  $t'$ to $t'_r$, and, subsequently, along the real time axis from  the real ionization time $t'_r$  to the final time $t \rightarrow \infty$ \cite{Popruzhenko2008a,Yan2012,Torlina2012,Torlina2013}. The first and second arms of the contour are associated with under-the-barrier dynamics and continuum propagation, respectively.

In the under-the-barrier part of the contour, we kept the momentum constant and equal to  $\mathbf{p}_0$ 
\cite{Maxwell2018b,Faria2020}. 
This reduces Eq.~(\ref{eq:tunncc}) to
\begin{equation}
\frac{\left[\mathbf{p}_0+\mathbf{A}(t')\right]^2}{2}=-I_p.
\label{eq:tunnccapprox}
\end{equation}
In the continuum propagation, the action can be simplified 
according to the procedure discussed in  \cite{Lai2015,Maxwell2017,Lein2016,Maxwell2020}.

The CQSFA transition amplitude obtained using the saddle-point approximation reads
\begin{equation}
\label{eq:MpPathSaddle}
M(\mathbf{p}_f)\propto-i \lim_{t\rightarrow \infty } \sum_{s}\bigg\{\det \bigg[  \frac{\partial\mathbf{p}_s(t)}{\partial \mathbf{r}_s(t_s)} \bigg] \bigg\}^{-1/2} \hspace*{-0.6cm}
\mathcal{C}(t_s) e^{i
	S(\mathbf{\tilde{p}}_s,\textbf{r}_s,t,t_s)} ,
\end{equation}where $\mathbf{p}_s$ and $\mathbf{r}_s$ and  $t_s$ are determined by Eqs.~(\ref{eq:p-spe}),(\ref{eq:r-spe}) and (\ref{eq:tunnccapprox}), respectively. The term in brackets gives the stability of a specific trajectory, and 
\begin{equation}
\label{eq:Prefactor}
\mathcal{C}(t_s)=\sqrt{\frac{2 \pi i}{\partial^{2}	S(\mathbf{\tilde{p}}_s,\textbf{r}_s,t,t_s) / \partial t^{2}_{s}}}\langle \mathbf{p}+\mathbf{A}(t_s)|H_I(t_s)|\psi_{0}\rangle.
\end{equation}
In practice, we use $\partial
\mathbf{p}_s(t)/\partial \mathbf{p}_s(t_s)$ instead of the stability factor in Eq.~(\ref{eq:MpPathSaddle}). This may be obtained employing a Legendre transformation and will not affect the action if the electron starts from the origin.  The continuum propagation is performed by solving the inverse problem, i.e., given a final momentum $\mathbf{p}_f(t)$, we find the initial momentum $p_0$ at the tunnel exit. This is defined as $\mathbf{r}_0(t'_r)$, where \begin{equation}
\mathbf{r}_0(\tau)=\int_{t'}^{\tau}(\mathbf{p}_0+\mathbf{A}(\tau'))d\tau'.
\label{eq:tunneltrajectory}
\end{equation}
A further approximation used here is to take a real tunnel exit 
\begin{equation}\label{eq:exitreal}
z_0=\mathrm{Re}[r_{0z}(t'_r)],
\end{equation}
where $r_{0z}$ indicates the component of the tunnel trajectory along the field polarization direction. 
This will simplify the problem and lead to real trajectories in the continuum. For discussions of how to solve the full complex problem see \cite{Pisanty2016,Maxwell2018b}.

The term $\langle \mathbf{p}+\mathbf{A}(t_s)|H_I(t_s)|\psi_{0}\rangle$ contains the influence of the initial bound state, i.e., its geometry, and will be important for the present work. Explicitly, 
\begin{equation}
\langle \mathbf{p}|H_I(t_s)|\psi_{0}\rangle=
\int d^3\mathbf{r}
\frac{\exp(-i \mathbf{p}\cdot \mathbf{r})}{(2 \pi)^{3/2}} \mathbf{r}\cdot \mathbf{E}(t_s)\psi_0(\mathbf{r}),
\end{equation}
where 
\begin{equation}
\psi_0(\mathbf{r})=\sum_{\alpha}
\frac{c_{\alpha}}{\sqrt{2}(2\pi)^{3/2}}
\phi_{\alpha}(\mathbf{r}),
\end{equation}
and $\phi_{\alpha}(\mathbf{r})$ is given by a Gaussian basis set
\begin{equation}
\phi_{\alpha}(\mathbf{r})=\sum_{v}b_v x^{\beta_x} y^{\beta_y} z^{\beta_z}
\exp\left( -\xi_v(x^2+y^2+z^2) \right).
\end{equation}
For details see \cite{Faria2010}.

In the CQSFA, one may identify four main types of orbits: 
\begin{enumerate}
	\item \textit{Type 1 orbits} behave like short, direct SFA trajectories that are slowed down by the Coulomb potential. They leave in the direction of the detector and their transverse momentum does not change direction.
	\item \textit{Type 2 orbits} are field-dressed Kepler hyperbolae whose start times are displaced in half a cycle from those in type 1 orbits. They start on the ``wrong" side and are turned towards the detector. They are lightly deflected by the potential and can be loosely associated with the long SFA orbits. Their transverse momentum does not change its sign during continuum propagation.
	\item\textit{ Type 3 orbits} are also field-dressed Kepler hyperbolae starting in the same half cycle as type 2 orbits, but interact more strongly with the core. Their transverse momentum changes direction during continuum propagation.  They  have no counterpart in the SFA as they depend on the interplay between the driving field and the binding potential and are not embedded in a Born-type method. In previous work, we have approximated orbit 3 within an analytic piecewise model in which rescattering is incorporated. However, mimicking the effect of orbit 3 and accurately reproducing the spider was only possible by including the acceleration caused by the Coulomb potential during the continuum propagation and the Coulomb phase \cite{Maxwell2017a}. This is a substantially different scenario than that encountered with the SFA, be it direct or rescattered, or in approaches focused on short-range potentials \cite{Zhou2016}. In the limit of vanishing residual potential for the continuum propagation, orbits 2 and 3 become degenerate and tend to the long direct SFA orbit \cite{Lai2015}. 
	
	\item \textit{Type 4 orbits} start in the same half cycle as type 1 orbits, but go around the core before reaching the continuum. Their behaviour is close to that of SFA rescattered trajectories, and they may reach energies of around $10 U_p$. However, one should notice that the process they undergo is different and their propagation in the continuum is not restricted to the field-polarisation axis. 
\end{enumerate} 

One should note that the energies of orbits 2 and 3 may go beyond the direct ATI cutoff of $2U_p$, due to the influence of the residual potential. This contributes to blur the distinction between direct and rescattered, as mentioned above. A limitation of the CQSFA is that it does not properly account for processes involving excited bound states. The basis chosen in Eq.~(\ref{eq:Mpp}) by using the closure relation in $\tilde{\mathbf{p}}_0$ works well for continuum states but does not reproduce excitation processes accurately. Still, it does take into account highly excited states which are strongly coupled with the field and behave similarly to the continuum. The four types of orbits stated above were first identified in \cite{Yan2010}.

\section{Results}
\label{sec:results}

\label{sub:spectra}

In the following, we will analyse how different polarisations in the initial bound states influence angle-integrated photoelectron spectra and photoelectron momentum distributions, with emphasis on several counterintuitive features. We used a
bandwidth-limited mid-IR pulse of intensity $I= 3.2 \times
10^{13}$ W/cm\textsuperscript{2}, frequency $\omega= 0.775$
eV, that corresponds to a wavelength $\lambda=1600$ nm, linear polarisation along the z direction, with a ponderomotive energy ($U_p$)
of $7.65$~eV. Unless otherwise stated, the binding energy used for this initial
state is $3.286$ eV ($I_p= 0.1238$ a.u.).  For this parameter range, we have verified that the electron in the ground state orbital 1s is not appreciably affected by the
IR pulse in the TDSE solvers. 

For the B-Spline ADC method and the SFA, the pulse used is described by a cosine
squared envelope for the vector potential
\begin{equation}
\mathbf{A}(t)=-\frac{A_{0}}{\omega}\cos^{2}\left(\frac{\pi t}{T}\right)\sin(\omega t)\hat{e}_z,\;\{-T/2\leq t\leq T/2\}
\label{eq:LaserPulse}
\end{equation}
of amplitude $A_0$ and polarization vector $\hat{e}_z$, and with a pulse duration $T$ set to 10 cycles (ca. 50 fs).

The calculations
have been performed using a linear B-spline knot sequence \cite{Ruberti2014} with
a radial box radius of R\textsubscript{max} = 500 a.u.
and~\emph{N\textsubscript{b}}~=~625 radial B-splines. The maximum
angular momentum employed in the monocentric expansion of Eq.~(\ref{eq:BSpline1})
was~\emph{l}\textsubscript{max}~=~30. Convergence of the results with
respect to the basis set parameters has been checked.
A complex absorbing potential (CAP) with starting radius
R\textsubscript{CAP} = 400 a.u. has been used to absorb the wavefunction
and avoid its reflections from the grid boundary. Each and every
B-spline ADC(1) energy-dependent photoelectron spectra presented here
were calculated by coupling the TD B-spline ADC method to the t-surff
technique.

For the CQSFA, for simplicity we consider a linearly polarized monochromatic wave, whose vector potential is described by
\begin{equation}
\mathbf{A}(t)=A_0\cos{\omega t}\hat{e}_z.
\end{equation}
This is a good approximation for long enough pulses. In practice, we take the field to be over 20 cycles long, and include ionization events within up to four laser cycles, that is, the  range of ionization times $t'$ lies within $0 \leq t' \leq 8 \pi/ \omega$. A restricted temporal unit cell is necessary for practical purposes, but may introduce some boundary effects. For a complete discussion see \cite{Stanford2021}. For Qprop, we have employed a four-cycle flat-top pulse with additional linear half-cycle turn on and off for the comparison to the CQSFA and SFA models. In the comparison of Qprop with the B-spline ADC model, we have matched the appropriate pulse of $\sin^2$, whose vector potential is given by Eq.~(\ref{eq:LaserPulse}).  

\subsection{ATI Spectra}
\subsubsection{One and two-electron Schr\"odinger solvers}
\begin{figure}
	\begin{center}
		\includegraphics[width=0.75\textwidth]{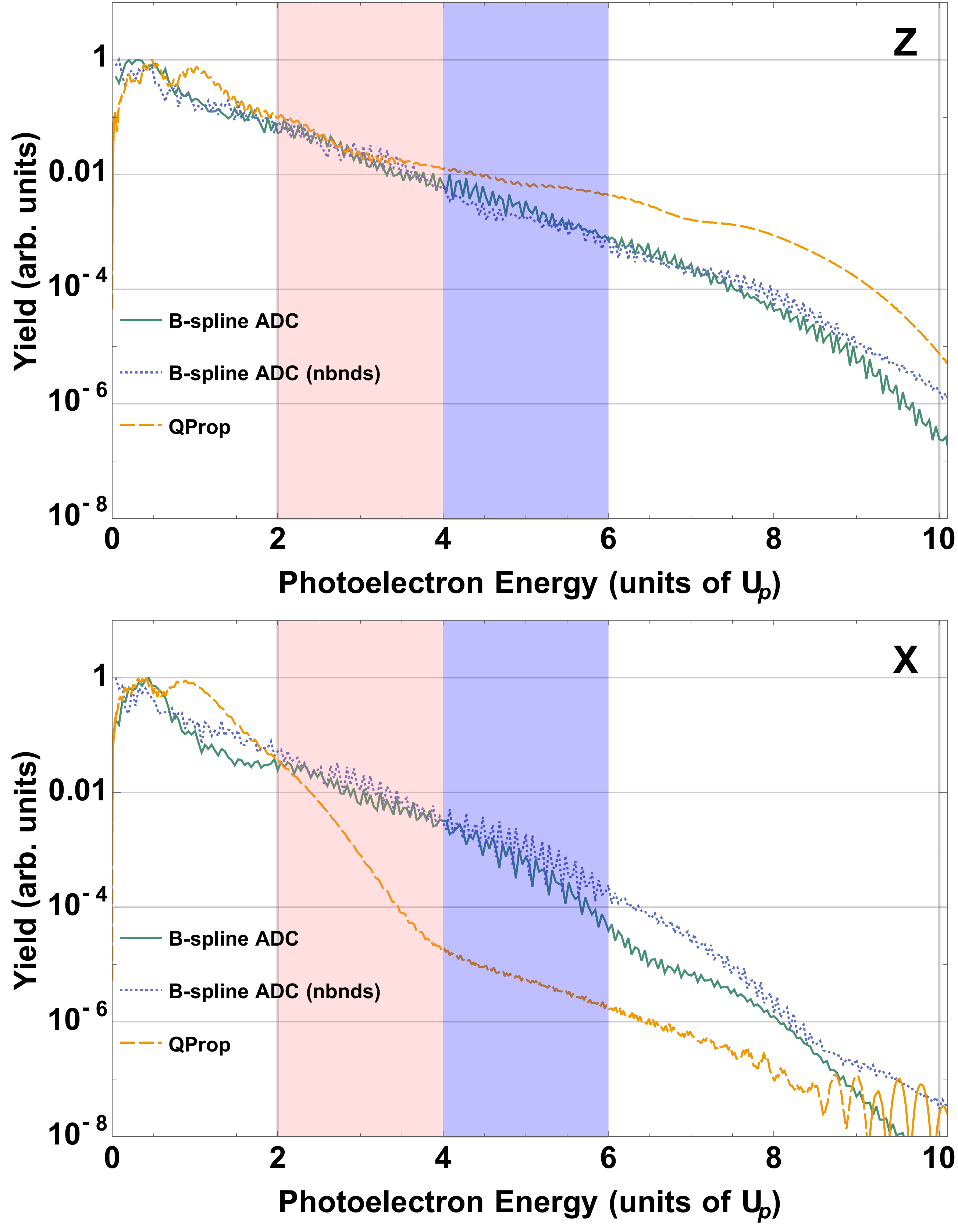}
		\caption{Angle-integrated photoelectron spectra computed for excited helium ($I_p = 0.1238$ a.u.) with numerical solutions of the time-dependent Schr\"odinger equation, for pulses with peak intensity $I_0 = 3.2 \times 10^{13}$ W/cm$^2$, ($U_p = 7.65$~eV ), and wavelength $\lambda =1600$~nm. The orange (dashed) curve gives the one-electron computation, performed with the free software Qprop \cite{qprop2,qprop3}, while the green (solid) and the blue (dotted) curves have been computed with the diagrammatic B-Spline ADC method including and excluding the bound-state transition $2^1P - 2^1S$, respectively. For the Qprop calculation a 4-cycle $\sin^2$ pulse was used, 
			while for the B-Spline ADC a 10-cycle $\sin^2$ pulse was taken and pulse length of 53 fs. The upper and lower panels have been calculated for $z$ (parallel) and $x$ (perpendicular) orientation of the initial $2p$ bound states of helium with regard to the driving-field polarization.  The blue curve in the lower panel has been computed for parallel polarization, but removing the bound-state transitions, the most important among them being $2^1P - 2^1S$. Each curve has been normalized to its largest value to facilitate a qualitative comparison. 
			The pink and violet shaded areas highlight the photoelectron energy ranges $2U_p\leq E_k \leq 4U_p$ and $4U_p\leq E_k \leq 6U_p$, respectively. }
		\label{fig:tdse}
	\end{center}
\end{figure}

We will start by discussing the angle-integrated ATI spectra obtained from the numerical solutions of the TDSE employed in this work, calculated with the diagrammatic B-Spline ADC method and with the freely available one-electron Schr\"odinger solver Qprop \cite{qprop2,qprop3}. These results are presented in Fig.~\ref{fig:tdse}, for the parallel and perpendicular initial-state configurations (upper and lower panel, respectively). For the parallel configuration, the spectra obtained with the B-Spline ADC method exhibits a long ramp-like structure extending from near the ionization threshold up to the high photoelectron energy of $10U_p$, while the Qprop computations exhibit a low-energy region followed by a ramp-like decrease from $2U_p$ to $4U_p$ (see pink shaded region), with a flat plateau up to $10U_p$. The high-energy suppression, in the case of TD B-Spline ADC, is caused by the “potential” that the excited/ionised electron experiences, which has contributions from both the Coulomb and exchange terms calculated with the Hartree-Fock orbitals of the helium atom. These exchange terms effectively repel the rescattered electron if it gets too close to the core, and thus render rescattering less effective. This influence increases with the photoelectron final energy, as higher-energy electrons must approach the core more closely. 
These ramp-like features  are not altered significantly if the bound-state transition $2^1P - 2^1S$ is removed from the B-Spline ADC computation (see blue and green curves).

On the other hand, if the initial $2p$ state is aligned perpendicular to the driving-field polarization (lower panel in Fig.~\ref{fig:tdse}), the computations lead to strikingly different results. For clarity, we have included the parallel-polarized spectrum without the $2^1P - 2^1S$ transition. One should note that, due to selection rules, this transition is forbidden in the perpendicular configuration. Up to the photoelectron energy of roughly  $2.5U_p$, there are at most subtle differences, with the Qprop spectrum being slightly larger than the B-Spline ADC spectra. Subsequently, one sees a steep ramp-like suppression for the (one-electron) Qprop computation up to $4U_p$ succeeded by a second ramp of much gentler slope. The B-Spline ADC spectra are significantly less suppressed for energies higher than $2U_p$, regardless of configuration. Thus, in contrast to the parallel-aligned case, electron exchange  appears to enhance rescattering.  The results suggest that exchange  effects, which are included in the B-spline ADC simulation, lead to a decrease of the dependence of the yield on the direction of the driving field with respect to the initial atomic polarization, especially in the (approximately) $2U_p-6U_p$ energy region. 
Interestingly, for the perpendicularly polarized initial state the B-Spline ADC computation exhibits multiple ramps instead of the monotonic behavior identified for parallel alignment. Thereby, one may identify two ramps, for photoelectron energy ranges $2U_p\leq E_k \leq 4U_p$ and $4U_p\leq E_k \leq 6U_p$, respectively.  Furthermore, in comparison to the parallel-aligned case (blue curve), the B-Spline ADC spectrum computed for the perpendicular configuration (green curve) exhibits a decrease of one order of magnitude for energies higher than $4U_p$ even in the absence of the  $2^1P - 2^1S$ transition. Still, this behaviour occurs for much larger energies than those in the Qprop spectrum. The remarkably different behaviour from Qprop suggests a multielectron character for the second ramp as well. 

\subsubsection{Comparison with orbit-based methods}
\begin{figure}
	\begin{center}
		\includegraphics[width=0.7\textwidth]{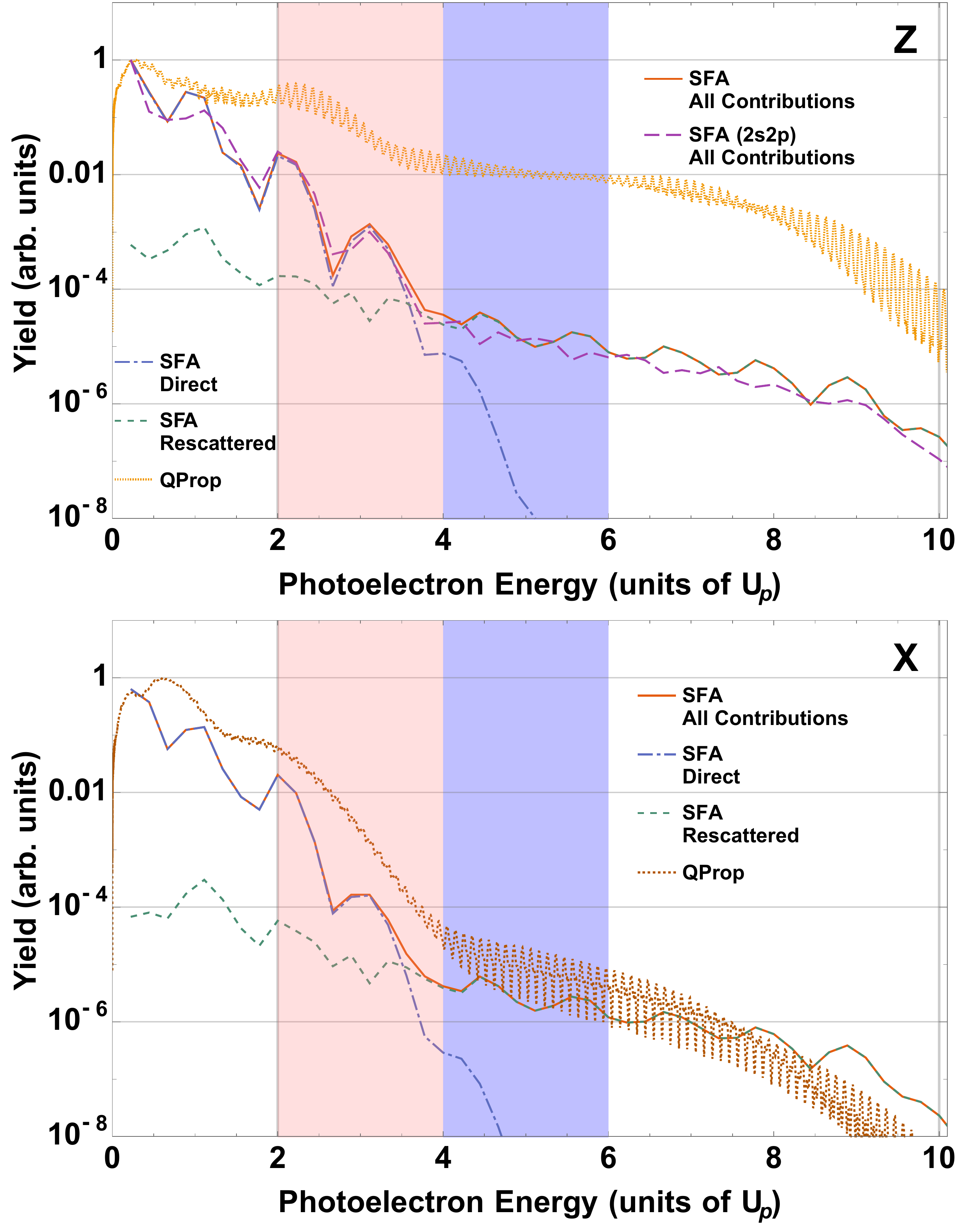}
		\caption{Angle-integrated photoelectron spectra computed for excited helium ($I_p = 0.1238$~a.u.) using Qprop (orange and brown dotted curves for parallel and perpendicular cases,  respectively) and the strong-field approximation (SFA) (remaining curves), for the same peak-field intensities and wavelengths as in the previous figures. For the Qprop calculation a flat-top four cycle pulse with an additional half a cycle turn on and off was used, while for the SFA we have taken only a single cycle. Each curve has been normalized to its largest value to facilitate a qualitative comparison. For the perpendicular configuration, the SFA spectrum is one order of magnitude smaller than its parallel counterpart in the direct region (energies smaller than $2U_p$ and the rescattered plateau is two orders of magnitude smaller. The top panel (parallel alignment) also shows all scattering (Direct and Rescattered)  contributions in the SFA for the bound coupled state, $2s2p$ (purple wide dashed curve).  The pink and violet shaded areas highlight the photoelectron energy ranges $2U_p\leq E_k \leq 4U_p$ and $4U_p \leq E_k \leq 6U_p$, respectively.  }
		\label{fig:QpropSFA}
	\end{center}
\end{figure}

For that reason, we will now compare the outcome of Qprop with that of semi-analytical, orbit-based methods. We will start by discussing the spectra obtained with the strong-field approximation (SFA), which is shown in Fig.~\ref{fig:QpropSFA}. In all cases, the energy regions in the SFA spectra can be clearly associated with direct and rescattered ATI, with a distinct cutoff at photoelectron energy $2U_p$ followed by long plateaus up to the rescattered ATI cutoff of $10U_p$. The direct contributions dominate up to approximately $4U_p$, with the plateau prevailing for higher energies. For perpendicular alignment, the plateau is much more suppressed than in the parallel-aligned case. This is consistent with the picture put across by the SFA, that rescattering occurs for a small angular range around the polarization axis. Because there is not much probability density with which the returning electronic wave packet can overlap, the resulting signal will be suppressed. This adds up to an overall suppression that occurs for similar reasons at the instant of ionization. Since tunneling is highly directional and occurs predominantly along and in the vicinity of the polarization axis, the overall signal will be around two orders of magnitude weaker in the perpendicular case. We have, however, normalized the yield in the lower panels of the figure in order to perform a qualitative comparison. 

In comparison with the Qprop calculation, the SFA plateau is strongly suppressed for parallel orientation, although they follow the same trend (see upper panel in Fig.~\ref{fig:QpropSFA}). In fact, it is noteworthy that there are only two orders of magnitude difference between the direct and rescattered signal in the Qprop case, while for the SFA this difference amounts to four orders of magnitude. Moreover, the behavior in the direct region is much flatter than for the SFA, with a ramp for photoelectron energies  between $2U_p$ and $4U_p$ (see pink shaded region). Interestingly, including the $2^1P - 2^1S$ bound-state transition in the SFA as discussed in Sec.~\ref{subsec:Bound-state-transitions} leads to subtle changes in the plateau, but does not alter its overall shape or intensity. 

For perpendicular-aligned initial states (lower panel in Fig.~\ref{fig:QpropSFA}), the Qprop and SFA computations are much more similar, with a steep decay in several orders of magnitude for $2U_p\leq E_k \leq 4U_p$ (dubbed `mid-energy ramp') followed by a plateau. Still, in that energy region the TDSE result always follows that of the SFA from above, and exhibit a ramp-like structure. One should note that this mid-energy ramp appears in a single-active electron setting. This is a key difference with regard to the behavior seen for the second ramp in the B-Spline ADC computations; see discussion of Fig.~\ref{fig:tdse}.  The mid-energy ramp is associated with hybrid orbits, which are missed by the SFA, and will be analyzed next.

\begin{figure}
	\begin{center}
		\includegraphics[width=0.7\textwidth]{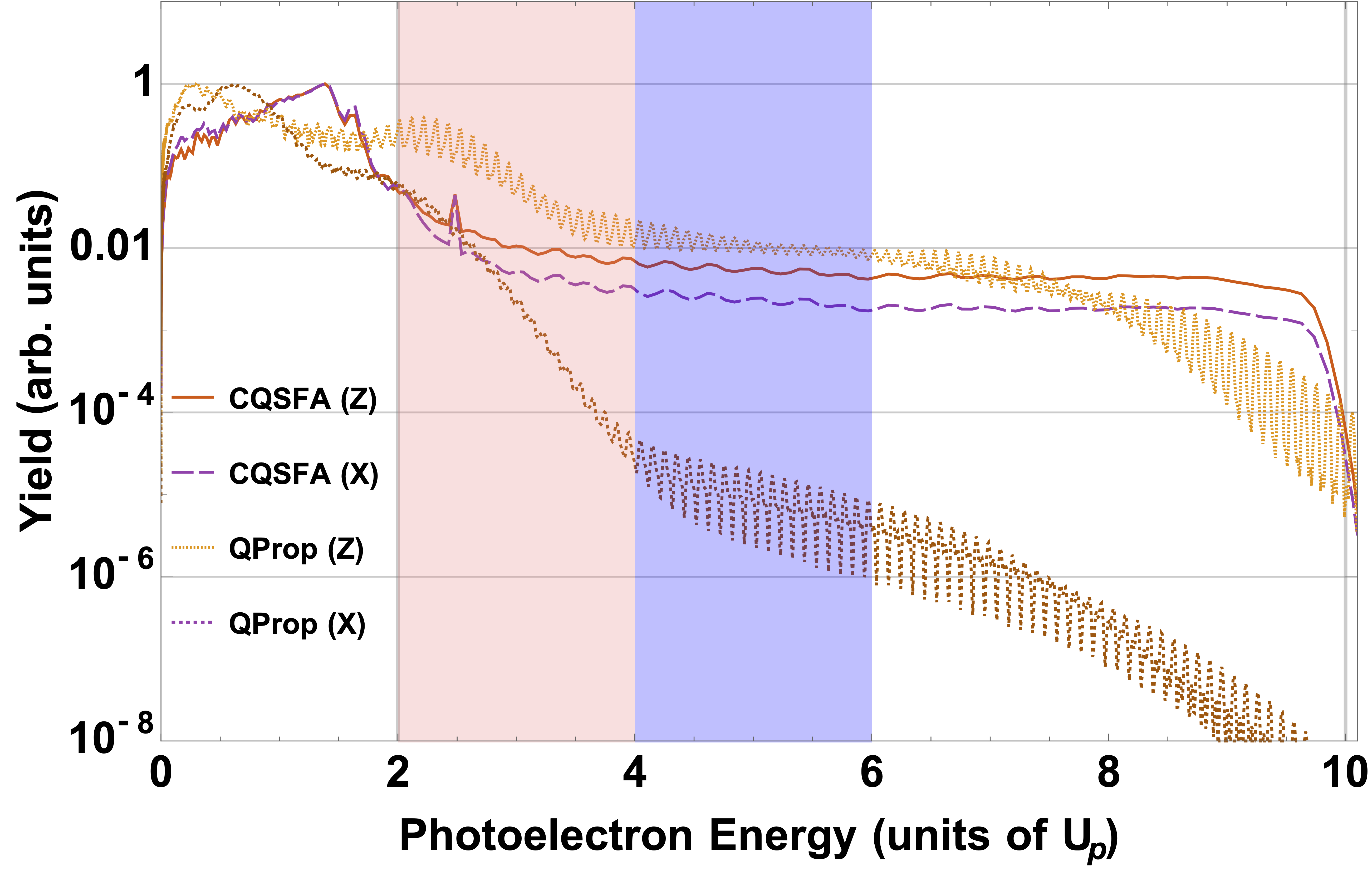}
		\caption{Angle-integrated photoelectron spectra computed for excited helium ($I_p = 0.1238$~a.u.) in initial $2p$ states oriented parallel (z) and perpendicular (x) to the laser-field polarization, using Qprop (dotted orange and brown lines) and the coulomb quantum-orbit strong-field approximation (CQSFA) (purple and red lines, dashed and solid respectively), for the same peak-field intensities and wavelengths as in Fig.~\ref{fig:tdse}. For the Qprop calculation, a flat-top four cycle pulse with an additional half a cycle turn on and off was used, while for the CQSFA, we have considered ionization events within one cycle of a 
			monochromatic field.  Each curve has been normalized to its largest value to facilitate a qualitative comparison. The pink and violet shaded areas highlight the photoelectron energy ranges $2U_p\leq E_k \leq 4U_p$ and $4U_p\leq E_k \leq 6U_p$, respectively.  }
		\label{fig:QpropCQSFA}
	\end{center}
\end{figure}

In Fig.~\ref{fig:QpropCQSFA}, we plot the spectra obtained with Qprop against those from the CQSFA. The spectra obtained with the CQSFA exhibit an almost constant, much higher plateau, whose onset occurs for energies immediately higher than $2U_p$. This is in stark contrast with the ramp-like behavior obtained in the previous cases.  Furthermore, a direct comparison of the parallel and perpendicular configurations shows a suppression of at least one order of magnitude for the plateau in the latter case. This is much less than that observed in all other methods, and it is also curious that the parallel and perpendicular spectra exhibit very similar shapes. This suggests that, for the parameter range employed, and in particular the ionization potential $I_p$ = 0.1238 a.u. of the excited states, the CQSFA orbits are behaving in an unusual way, which is much more dictated by the binding potential than by the laser field. 

\begin{figure}
	\begin{center}
		\includegraphics[width=\textwidth]{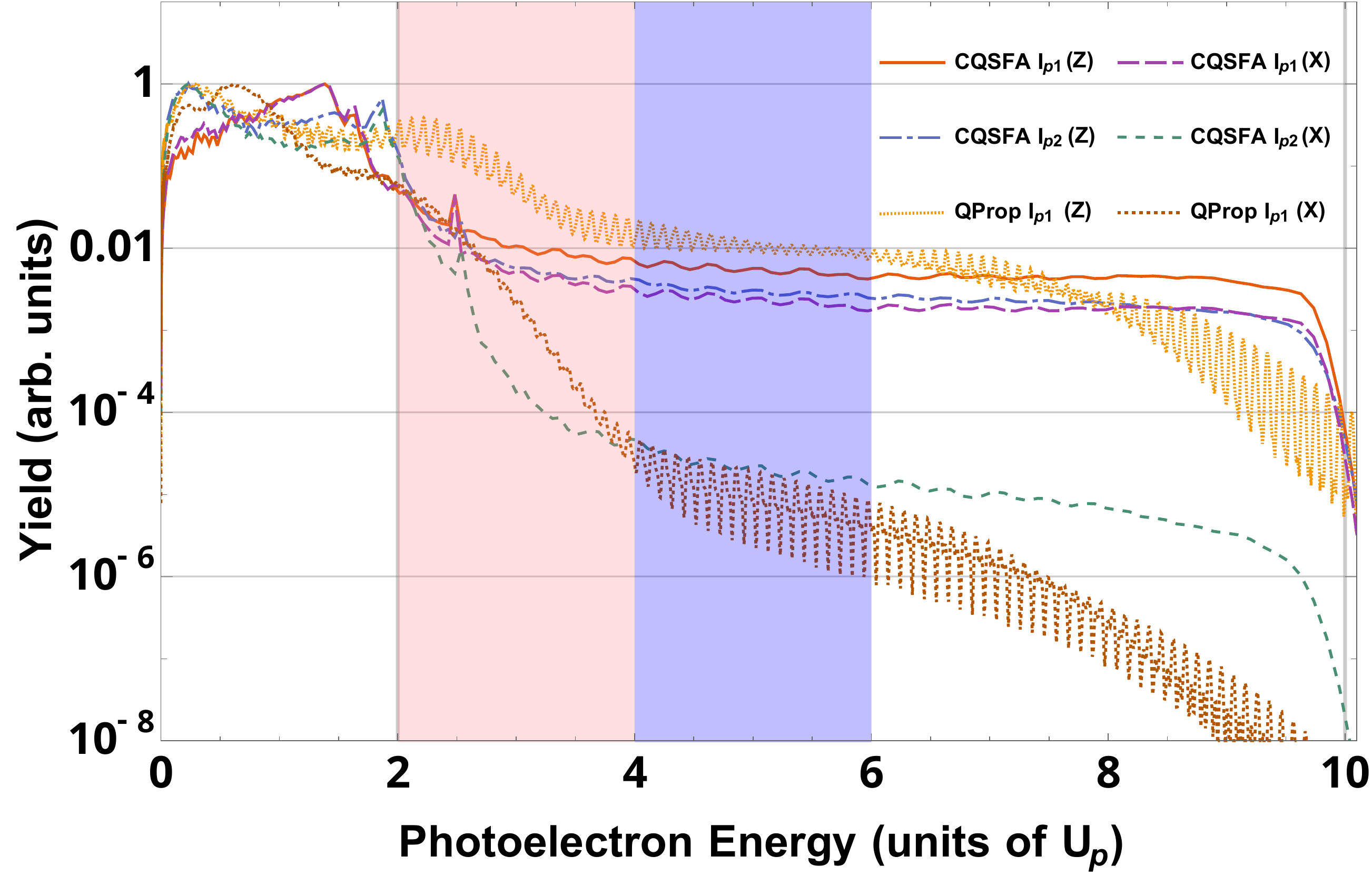}
		\caption{Angle-integrated photoelectron spectra computed for excited helium using Qprop (orange and brown dotted curves) and the Coulomb quantum-orbit strong-field approximation (CQSFA) (purple and red curves, wide dashed and solid respectively) computed for excited helium ($I_{p1} = 0.1238$~a.u.) using  the same parameters as in Fig.~\ref{fig:QpropCQSFA}, but with two additional (blue and green, dot-dahsed and small dashes respectively) curves computed using the CQSFA for a slightly larger ionization potential ($I_{p2} = 0.1750$~a.u.). Each curve has been normalized to its largest value to facilitate a qualitative comparison, except for those computed for the larger ionization potential $I_{p2}$. The pink and violet shaded areas highlight the photoelectron energy ranges $2U_p\leq E_k \leq 4U_p$ and $4U_p \leq E_k \leq 6U_p$, respectively.  }
		\label{fig:Ip}
	\end{center}
\end{figure}

A further investigation is presented in Fig.~\ref{fig:Ip}, in which, in addition to the previous curves, calculated for excited helium, we have included a CQSFA computation for a slightly larger ionization potential ($I_{p2}$ = 0.1750 a.u.) and $2p$ initial states parallel and perpendicular to the laser-field polarization. This is quite different from the behavior we have observed if the correct ionization potential ($I_{p1}$ = 0.1238 a.u.) is taken. Thereby, one can see that, for the perpendicular and parallel-oriented cases, the spectra  practically overlap up to energies up to $2U_p$. However, for the perpendicular case there is a long ramp in the photoelectron energy region  $2U_p\leq E_k \leq 4U_p$ followed by a strongly suppressed ATI plateau. This behavior is in better qualitative agreement with the Qprop outcome than that obtained for the correct ionization potential. Next, we will analyze these behaviors in more detail using photoelectron momentum distributions.    

\subsection{Photoelectron momentum distributions}
\label{sub:pmds}

\begin{figure}[H] 
	\begin{center}
		\includegraphics[width=\textwidth]{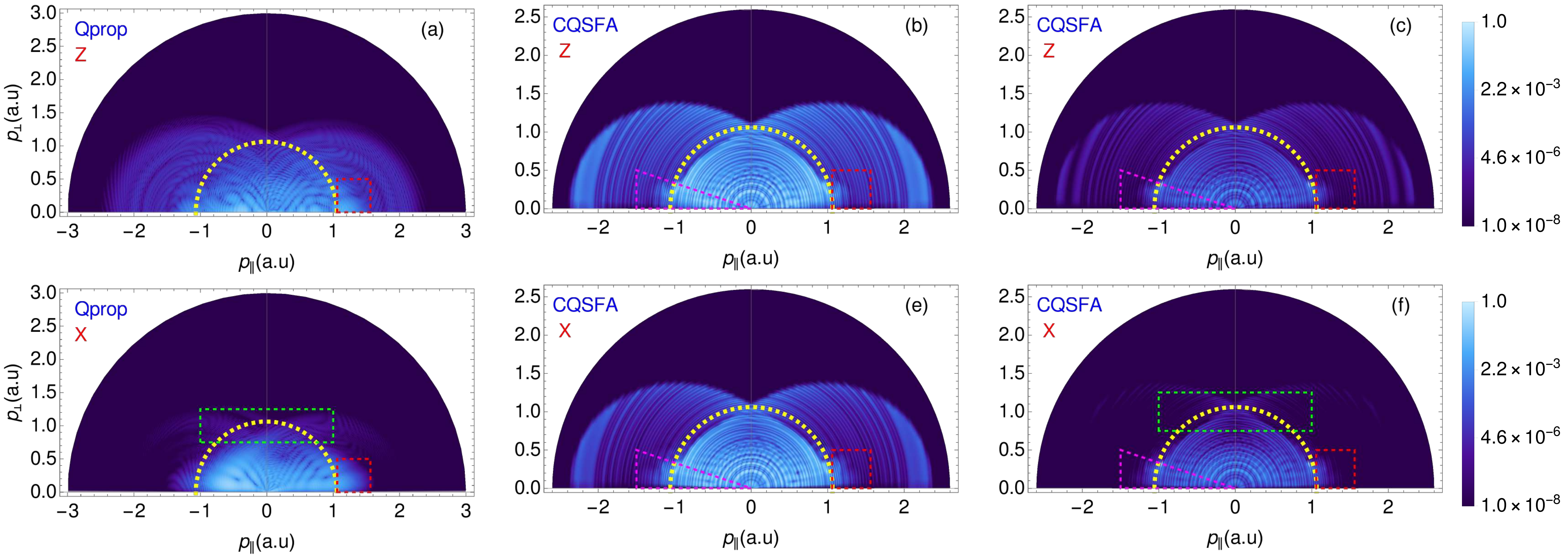}
		\caption{Photoelectron momentum distributions computed using the one-electron Schr\"odinger solver Qprop [panels (a) and (d)] and the Coulomb quantum-orbit strong-field approximation (CQSFA) [panels (b) and (e)] with all relevant orbit types for excited helium at the given $I_p$ = 0.1238 a.u. In panels (c) and (f), we have considered the CQSFA and a slightly larger ionization potential ($I_p$ = 0.175 a.u.)  as a test case. The upper and lower rows have been computed for initial bound states aligned parallel and perpendicular to the driving-field polarization, respectively.  For Qprop, we have used a four-cycle flat-top pulse with an additional half cycle turn on and off. For the CQSFA, we have approximated the driving field by a monochromatic wave.  In all cases, we considered a driving-field peak intensity $I_0 = 3.2\times10^{13}$~W/cm$^2$ ($U_p = 7.65$~eV), and wavelength $\lambda =1600$~nm. The thick circular dashed line indicates the direct ATI cutoff $2U_p$ and the yield in each panel has been normalized to its maximum value, but there are roughly between one and two orders of magnitude difference between the parallel and perpendicular configuration. 
			Furthermore, the yield for the perpendicular case is suppressed by a factor of $10^{2}$ for CQSFA and $10$ for Qprop, respectively. The green dashed rectangle on panels (d) and (f) indicate the suppressed rescattering ridge for a particular perpendicular momentum, whereas the dashed triangle shows off axis suppression in panels (b), (c), (e) and (f). Finally, the red squares seen on all panels show the spilling beyond the $2U_p$ cutoff as mentioned above.  }
		\label{fig:AllOrbitPAD}
	\end{center}
\end{figure}

In the following, we will have a closer look at the electron dynamics by inspecting photoelectron momentum distributions (PMDs). For simplicity, we restrict ourselves to a one-electron scenario employing the CQSFA and using Qprop \cite{qprop2,qprop3} as a benchmark. These results are presented in  Fig.~\ref{fig:AllOrbitPAD}, for parallel and perpendicularly polarized helium (upper and lower rows, respectively) in a four-cycle trapezoidal pulse. There is an overall suppression of the signal in the perpendicular configuration, but in the present discussion we will focus on qualitative features and the physics behind them. All distributions exhibit well-defined ATI rings and several holographic structures, such as a spider-like pattern around the polarization axis, fan-shaped fringes close to the ionization threshold and spiral-like patterns perpendicular to the driving-field polarization. These structures have been discussed in previous publications \cite{Maxwell2017,Maxwell2017a,Maxwell2018,Maxwell2018b} and have been observed in numerical computations and experiments (for details see the recent review article \cite{Faria2020}). There are, however, some qualitative differences between the CQSFA and the TDSE computations. First, the holographic patterns in the TDSE outcome are asymmetric with regard to the $p_{\perp}$ axis and the ATI rings are less sharp. This asymmetry is very likely caused by a strong bound-state depletion, which is present in the TDSE case. Depletion will affect the probabilities associated with different ionization events such that the later events will be suppressed, and this will cause distortions in the patterns, making the distributions more similar to those from a few-cycle pulse. As such, we have verified that the initial excited-state population is quickly reduced for the parameter range employed (not shown). This effect has not been incorporated in the CQSFA. The CQSFA also has some asymmetry for a different reason. Despite using a monochromatic field, we must restrict the ionization events to a finite temporal range, which introduce some artificial asymmetry depending on how this range is chosen.  This can be eliminated by the incoherent averaging procedure described in \cite{Stanford2021}.

Another noteworthy aspect in the PMDs are superimposed circular regions of radius $10U_p$ centered at $(p_{\parallel},p_{\perp})=(\pm 2\sqrt{U_p},0)$, which overlap at $(p_{\parallel},p_{\perp})=(0,2\sqrt{U_p})$ along the perpendicular momentum axis. The boundaries of such regions determine a ridge, which is a well known indicator of rescattering. The rescattering ridge is present in the CQSFA computation, regardless of the orientation of the initial orbital [see middle panels in the figure], but is strongly suppressed for the TDSE in the case of perpendicular orientation [see lower left panel]. This suppression is not isotropic, and remnants of the ridge can be seen around the axis $p_{\perp}$ perpendicular to the driving-field polarization.  If, on the other hand, we consider a slightly larger ionization potential for the CQSFA, we also observe a very faint rescattering ridge for the perpendicular case (see right column in the figure). In addition to that, there are suppressions of the photoelectron signal around the $p_{\perp}$ axis, which are stronger for perpendicular alignment. Subtler features are a spilling of the spider-like structure beyond the direct $2U_p$ cutoff and a wedge-like suppression (shown on Fig.~\ref{fig:AllOrbitPAD} by the dashed square and triangle, respectively.) along the $p_{\parallel}$ axis for the perpendicular bound-state configuration in all approaches. 

\begin{figure}[h]
	\begin{center}
		\includegraphics[width=0.8\textwidth]{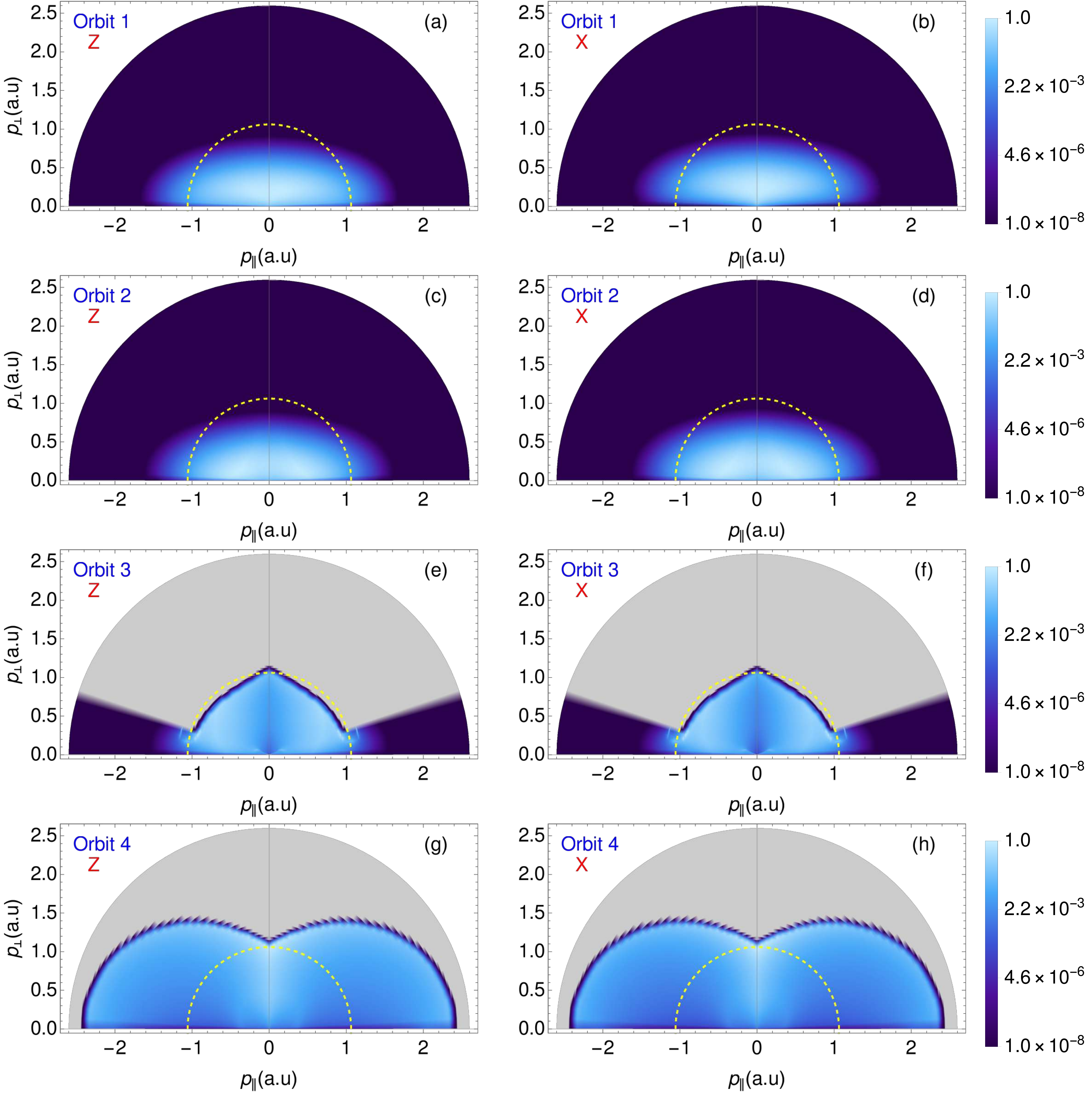}
		\caption{Single-orbit electron momentum distributions for excited helium ($I_p = 0.1238$~a.u.) using the same field parameters as in Fig.~\ref{fig:AllOrbitPAD} and a single cycle, for parallel and perpendicular alignment with regard to the laser-field polarization. From left to right, we present the contributions of orbit 1 [panels (a) and (b)], orbit 2 [panels (c) and (d)], orbit 3 [panels (e) and (f)] and orbit 4 [panels (g) and (h)], where the first column is the parallel configuration and the second is the perpendicular configuration. The thick circular dashed lines indicate the direct-ATI cutoff energy $2U_p$, and the yield on each panel has been normalized to its maximum value, but there is an overall suppression of over one order of magnitude for the perpendicular case. Orbits 3 and 4 are not defined in the grey regions displayed in panels (e), (f), (g) and (h). }
		\label{fig:SingleOrbitPAD}
	\end{center}
\end{figure}

We will next analyze the above features in terms of CQSFA orbits. With that aim in mind, in Fig.~\ref{fig:SingleOrbitPAD}, we plot PMDs obtained using individual CQSFA orbits. The plots illustrate the momentum region they occupy, and how they are influenced by the parallel and perpendicular configurations. 
We will commence by looking at the parameters in Figs.~\ref{fig:AllOrbitPAD}(b) and (e), which, apart from an overall suppression of roughly two orders of magnitude in the perpendicular case, look qualitatively similar.

Orbits 1 and 2 lead to approximately elliptical PMDs, whose major axis is along $p_{\parallel}$ and which are mostly constrained by the $2U_p$ direct ATI cutoff. This is expected as both orbits behave primarily as direct SFA orbits. Spilling beyond this cutoff occurs mostly along the driving-field polarization and is caused by the presence of the Coulomb potential.  For perpendicular orientation, there is a wedge-like suppression along the $p_{\parallel}$ axis for the contributions of orbit 1 [see Fig.~\ref{fig:SingleOrbitPAD}(b) in comparison with Fig.~\ref{fig:SingleOrbitPAD}(a)], whose angular range broadens as $|p_{\parallel}|$ increases. This effect can be understood as follows: if an electron propagates along orbit 1, its final and initial momentum will be similar, and, for high enough final momenta, close to the standard SFA \cite{Maxwell2017}. Thus, if the electron is freed along the field-polarization axis, it will continue to propagate along this direction. For small angles, its propagation direction will not change significantly. The $2p$ orbital is strongly directional, so that if it is oriented perpendicular to the laser-field polarization, tunneling along the $p_{\parallel}$ axis will be hindered.  The wedge-like shape comes from the fact that this is not a perfect mapping, and the final and initial momenta will only coincide for large values of $p_{\parallel}$. 
Still, the above discussion will approximately hold. In contrast, as orbit 2 is essentially a field-dressed hyperbola, the binding potential plays a stronger role and the transverse momentum mapping is non-trivial. In this case, the wedge-like feature is absent. 

\begin{figure}
	\centering
	\includegraphics[width=\textwidth]{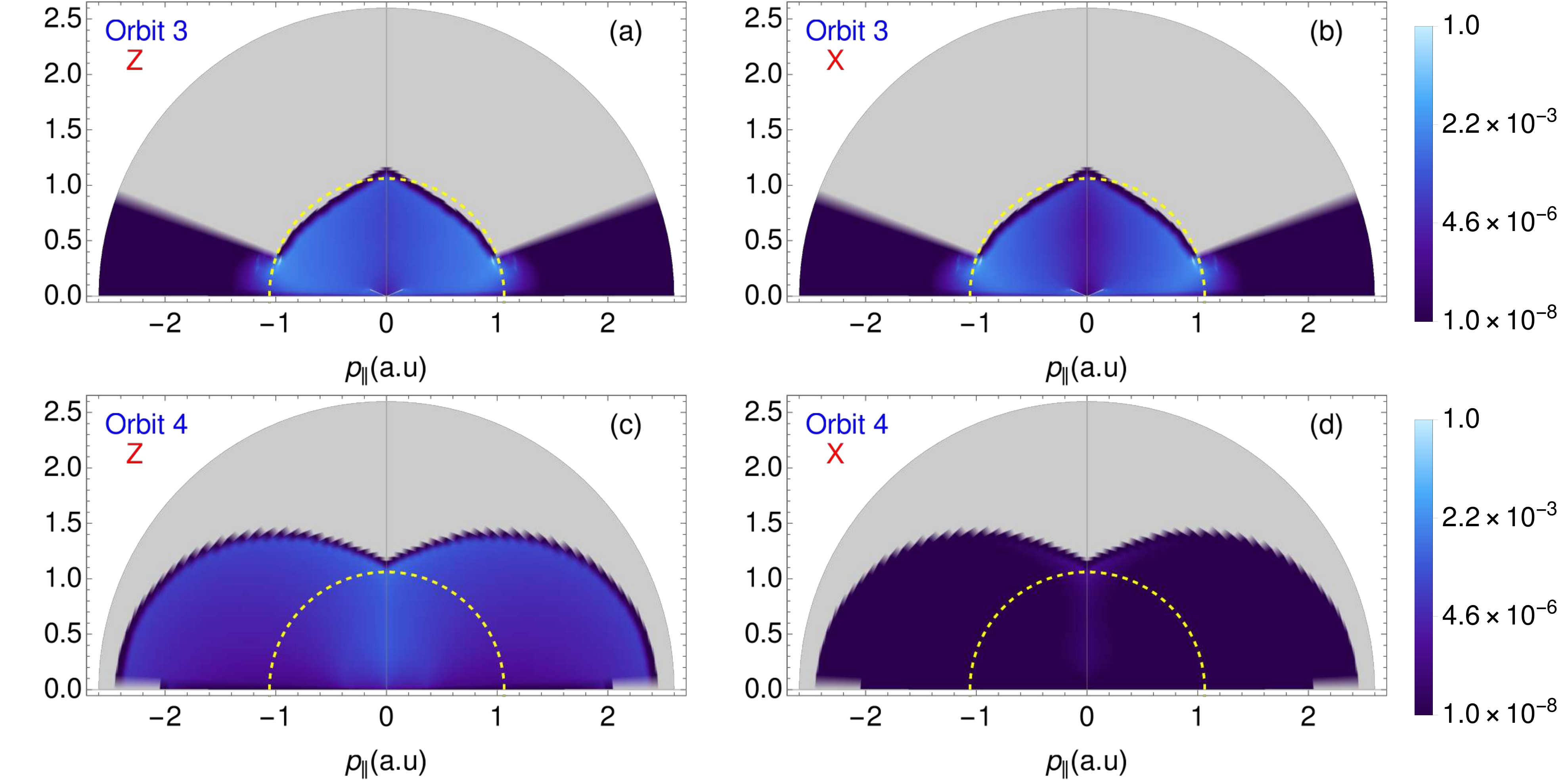}
	\caption{Single-orbit distributions calculated for orbits 3 [first row] and 4 [second row] using the CQSFA for the same field as in Fig.~\ref{fig:AllOrbitPAD} and a single cycle,  but considering a slightly higher ionization potential ($I_p = 0.175$~a.u.). The first columns panels (a) and (c) have been computed for parallel case. Whereas perpendicular alignment with regard to the driving-field polarization, is the second column, panels (b) and (d) and the  circular lines indicate the direct ATI cutoff at energy $2U_p$ across all panels.  The two upper panels have been normalized to their highest value, but there is a difference of one to two orders of magnitude between the parallel and perpendicular cases. In the lower panels the same scale was used. }
	\label{fig:Orbs34_highIp}
\end{figure}

The PMDs associated with orbits 3 and 4, shown in the bottom rows in Fig.~\ref{fig:SingleOrbitPAD}, occupy momentum regions within a broader angular range than the previous single-orbit distributions. This is due to them undergoing a stronger deflection and interaction with the core.  Previous studies have shown that orbit 3 has a hybrid character \cite{Maxwell2017a,Maxwell2018}, with no counterpart in the SFA, while orbit 4 behaves essentially like a rescattered orbit. The PMD associated with orbit 3 mainly occupies the momentum region defined by the overlapping rescattering ridges, with a slight spilling near the polarization axes. In both parallel and perpendicular configurations, there is a suppression along $p_{\parallel}=0$, which, however, is more pronounced in the latter case. 
Orbit 4 leads to a PMD with clear rescattering ridges for both parallel and perpendicular configurations and a caustic structure at the boundary. This indicates that, for the present parameter range, the dynamics associated with orbit 4 are not predominantly located along the polarization axis, as would be expected in an SFA treatment of rescattered trajectories. On the other hand, if we consider a slightly larger ionization potential, there is a much stronger suppression of the ridge for orbit 4 and of the signal around the $p_{\perp}$ axis for orbit 3.  This is illustrated in Fig.~\ref{fig:Orbs34_highIp}.

These are great examples of how rescattering may no longer be restricted to the polarization axis due to the Coulomb potential.  For the correct value of the ionization potential associated with the excited state, the CQSFA tunnel exit is located in a region for which the Coulomb potential is dominant. This suggests that orbit 4 will behave in a much less directional way than what one expects within the SFA framework. It will be able to probe orbitals oriented perpendicular to the driving-field polarization, and the rescattering ridge will be present. On the other hand, for higher ionization potentials, the tunnel exit will be located in the region for which the field is dominant. This will reduce the angular range for which rescattering may occur, and will cause a strong suppression for photoelectron energies higher than $2U_p$. 
\begin{figure}
	\centering
	\includegraphics[width=\textwidth]{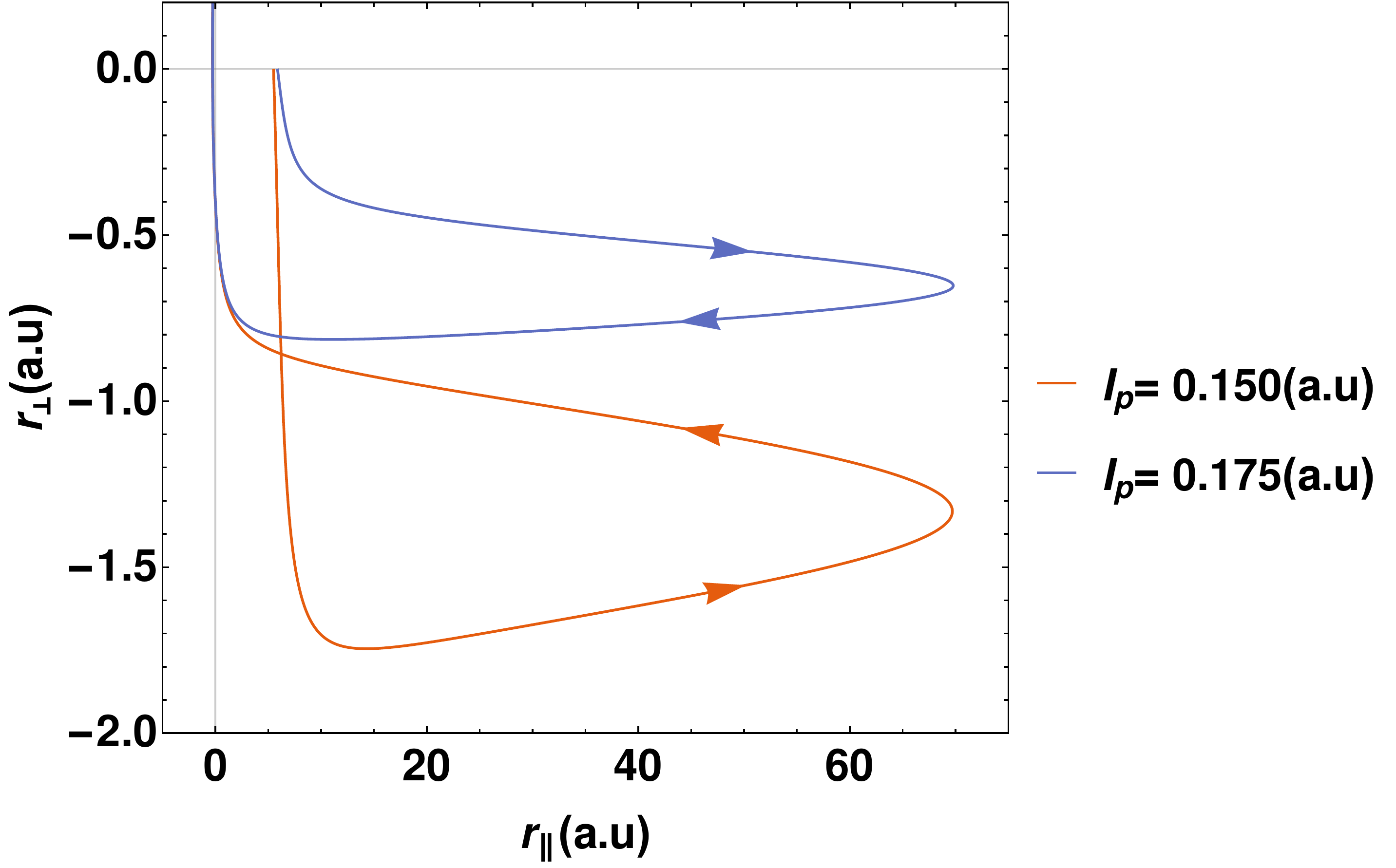}
	\caption{The behavior of orbit 4 for different ionization potentials ($I_p=0.150$~a.u. and $I_p=0.175$~a.u.), plotted over one cycle of the driving field and calculated for the same driving-field parameters as in Fig.~\ref{fig:SingleOrbitPAD}. The tunnel exit for the orange curve occurs at $5.46$~a.u in parallel direction, while for the blue curve it is located at $5.89$~a.u. The final momentum components associated with the orange and blue curves are $1.90$~a.u in the parallel and $0.53$~a.u. in the perpendicular direction.}
	\label{fig:orbitsIp}
\end{figure}

This issue is illustrated in Fig.~\ref{fig:orbitsIp}, which shows that this transition depends quite critically on the interplay between the ionization potential and the ponderomotive energy. For the ionization potential $I_p=0.15$ a.u., which is slightly larger than that of excited helium (orange curve), the binding potential dominates. This causes orbit 4 to leave in a perpendicular direction to the field, until approximately 2 a.u.  Subsequently, the orbit leaves and returns to the core region propagating in the expected direction, that is,  parallel to the driving-field polarization. Still, the large initial transverse momentum acquired by orbit 4 makes it much less directional than what is predicted by the SFA. This means that, for that parameter range, this orbit will also probe orbitals with perpendicular alignment. This explains the similar CQSFA spectra in the previous section, and the prominent rescattering ridges in Fig.~\ref{fig:SingleOrbitPAD}(h) and Fig.~\ref{fig:AllOrbitPAD}(e). In contrast, if the tunnel exit is located in a region for which the field is dominant (blue curve), orbit 4 will behave in a much more directional way, practically along the polarization axis. This will cause an overall suppression for initial bound states oriented perpendicular to the driving-field polarization. Nonetheless, there are still momentum components in the perpendicular direction and the orbit does not leave parallel to the field. This, together with the presence of intermediate orbits, leads to remnants of the rescattering ridge in the PMDs.

Therefore, even in a single-electron setting, the residual binding potential influences rescattering. It gives rise to orbits of hybrid character, and, in some instances, makes ionization and/or recollision deviate from the driving-field polarization axis. Hybrid orbits are absent in Born-type approaches such as the SFA, and will contribute for the anisotropic suppression in the rescattering ridge reported in this work. Ionization, and in some cases propagation,  perpendicular to the laser-polarization axis may occur and will partially hinder the suppression of the rescattered ATI signal. One should note, however, that the CQSFA results for excited helium strongly deviate from the TDSE computations. Interestingly, the TDSE agrees more with the CQSFA if a higher (artificial) ionization potential is taken in the CQSFA. 
\section{Conclusions}
\label{sec:conclusions}

In this work, we have probed excited helium in initial $2p$-state configurations parallel and perpendicular to a mid-IR, linearly polarized field. We use different numerical methods, namely the B-spline algebraic diagrammatic construction (ADC) \cite{Ruberti2014} and the one-electron Schr\"odinger solver Qprop \cite{qprop2,qprop3}, whose outcomes were compared to those of two semi-analytic methods: the Strong-Field Approximation (SFA) and the Coulomb quantum-orbit strong-field approximation (CQSFA). Our main goal was to understand the limitations of the rescattering model in its simplest form,  i.e., that dictated by the SFA, and what subtleties must be taken into consideration. Overall, it is noteworthy that different computations may lead to spectra whose shapes and intensities are quite distinct.

The SFA predicts a sharp decrease in the ATI signal for energies above the direct ATI cutoff $2U_p$ for all cases, and a strong suppression in the high-energy ATI plateau, which extends up to the energy of 10$U_p$, for the perpendicular initial state. This is due to tunnel ionization and recollision being located around the field polarization axis. This implies that both processes are suppressed due to the initial-state geometry. First, there will not be a substantial probability density to undergo tunneling and, upon the act of rescattering, there will not be a significant overlap of the returning wavepacket with the orthogonally oriented bound state. Furthermore, in the SFA no changes that may occur in the core and alter its geometry are incorporated. 
This sharp decrease is similar to that resulting from changing the driving-field polarization, instead of that of the initial bound state.  Any deviations from those patterns mean that the dynamics are more intricate, either because ionization or recollision may occur off axis, or because processes involving the core must be taken into consideration. 

The results presented in this paper reveal a subtler picture, with the  angle-integrated ATI spectra exhibiting ramp-like structures in some computations. This is most striking for the B-spline ADC computation, which goes beyond a single-active electron and has the least degree of physical approximations. The B-spline ADC spectra exhibit multiple ramps and a smooth transition from direct to rescattered ATI, both for parallel and perpendicular orientation. A ramp-like structure is expected if there are many superimposed processes and/or substantial depletion. This will tend to weaken effects caused by the initial bound-state geometry or suppress high-energy features by removing parts of the bound-state population via several channels. It could also be caused by mechanisms rendering rescattering less effective. 

It is noteworthy that the B-Spline ADC spectra behave in strikingly different ways from those obtained with the one-electron computations. The spectra from Qprop and  the analytical models exhibit a much more obvious transition from the direct to the rescattered regimes, while the B Spline ADC spectra present a continuous ramp. 
Removal of a specific bound-state excitation channel ($2^1P - 2^1S$) in the B-Spline ADC influences the spectra. However, the changes are too subtle to justify all the discrepancies. The current results suggest that in the parallel-aligned case the presence of electron exchange renders rescattering less effective by introducing repulsive effects, while for the perpendicular-aligned configuration multielectron effects enhance tunneling and low-energy rescattering. Because exchange is non-local, it renders both effects less directional, which decreases the suppression in the perpendicular case. Interestingly, a suppression of one order of magnitude between parallel and perpendicular polarised initial $p$ states has also been observed for ATI spectra computed for argon using the time-dependent density functional theory (TDDFT) \cite{Tong2015}. At the ADC(1) level at which our simulations were done, we are close in accuracy to the TDDFT, and both computations incorporates electron exchange.  This suggests that rescattering in ATI is a delicate effect that probes the fine details of the core structure and core-photoelectron interaction including its non-local (exchange) component.  The influence of electron exchange will also be energy dependent. The closer to the core  the rescattering electron gets, the higher the photoelectron energy will be. This also means that exchange effects will become increasingly important, leading to a ramp.  

Still, even for single-electron computations, the spectra exhibit quite different features, depending on what is included or left out. For instance, in the parallel-aligned case, the SFA spectrum shows a much stronger drop in signal for the plateau than that of Qprop. In the perpendicular-aligned scenario, the Qprop outcome follows the SFA spectrum from above in the direct region, and decays sharply after the direct ATI cutoff. In contrast, in the CQSFA the plateau is overestimated unless an artificially large ionization potential is taken. Moreover, it varies much less dramatically with the initial states' orientation than all other methods considered in this work. 

The features mentioned above can be understood in greater depth by looking at angle-resolved photoelectron momentum distributions (PMDs) and by analyzing the electron orbits involved. This analysis has been performed for the CQSFA and Qprop. The agreement is good for the main features, such as the rescattering ridge and a suppression near the polarization axis that occurs for the perpendicular configuration. However, the holographic patterns are much more irregular for Qprop than for the CQSFA. Furthermore, in the perpendicular configuration the rescattered ATI signal is much more suppressed for Qprop. This includes most of the signal beyond the energy of $2U_p$ and the rescattering ridge. A better agreement is obtained if the ionization potential is artificially increased in the CQSFA computation. 

The irregularities in the Qprop holographic patterns are explained by the huge amount of bound-state depletion, which favours certain events over others and compromises the contrast in the quantum interference patterns. Furthermore, there may be other mechanisms through which the electron is freed in the continuum, such as over-the-barrier ionization and coupling with highly excited states. These features are not incorporated in the CQSFA.  The over-enhancement of the rescattered contributions in the perpendicular case is due to the Coulomb potential being dominant upon ionization, for the CQSFA orbit type that leads to the rescattering ridge. This introduces a large perpendicular momentum component in the orbits, so that they initially tunnel in a perpendicular direction and may return farther away from the polarization axis.

Further studies indicate that a tunnel exit in a region for which the field dominates, which can be achieved with a slightly larger ionization potential, will alter this behavior. In this case, the orbit will be more localized around the polarization axis, so that tunneling and rescattering will be suppressed for perpendicular configurations. Interestingly, this suppression is closer to the behavior observed for the TDSE computation. The TDSE outcome will be influenced by ac Stark shifts, additional ionization channels, bound-state broadening and over-the-barrier ionization, which are not included in the CQSFA. Support for this assumption is provided by further TDSE computations using weaker binding energies, which exhibit a rescattering ridge for the perpendicular-aligned initial states (not shown). Some combination of these effects likely leads to the rescattering behaviour observed in the CQSFA when a larger ionization potential is used.
Discussions of how to include Stark shifts in the SFA are provided in \cite{Dimitrovski2010,Etches2010}. For a proposal of how to measure Stark shifts in excited helium, see \cite{He2011, Chini2012}. 

If, on the one hand, this suggests that the CQSFA may need modifications for highly excited, loosely bound states, on the other hand this is a nice illustration of how the interplay of the binding potential and the external laser field may lead to counterintuitive behaviors and move rescattering or ionization away from the driving-field direction. This, together with the core dynamics effects observed for the B-spline ADC spectra, shows that there may be other possibilities for imaging than those based exclusively on the field exploring bound-state geometry or driving-field shapes.    
Finally, one should note that preparing targets in excited and oriented states is within experimental reach. For instance, one may use the 
 Magneto-optical trap recoil ion momentum spectroscopy (MOTRIMS) technique (see, e.g. \cite{Blieck2008,Fischer2012} in the context of electron collisions and  \cite{Yuan2020} for multiply ionized Rubidium in a strong elliptically polarized field). Moreover,  one may prepare the  2 $^1$P state of helium either by resonant excitation from the metastable 2 $^1$S state \cite{Williams1968,Fry1969}, or by a direct extraction from the ground state \cite{Larsson1995,Gisselbrecht1999}. The polarization of the excited state can be controlled as described in \cite{Johansson2003}, and the lifetime of the 2$^1$P state is several orders of magnitude larger than the timescales involved in the present work. Usual excitation results in a superposition of the ground and excited state helium. However, for the parameters employed in this work, the ground state is too strongly bound to result in any appreciable ATI. Therefore, we are probing the excited-state population and whatever population stays in the ground state is simply lost for us. 

\section{Acknowledgements}
We would like to thank Dr Bridgette Cooper for her insights and tutorials in using GAMESS and Dr Emilio Pisanty for his knowledge on colour schemes to enhance plots without the loss of key features. ASM acknowledges grant EP/P510270/1 and CFMF grant No.\ EP/J019143/1, both funded by the UK Engineering and Physical Sciences Research Council (EPSRC). ASM and MC acknowledge support from ERC AdG NOQIA, Agencia Estatal de Investigación (``Severo Ochoa'' Center of Excellence CEX\allowbreak{}2019-000910-S, Plan National FIDEUA PID2019-106901\allowbreak{}GB-I00/10.13039 / 501100011\allowbreak{}033, FPI), Fundació Privada Cellex, Fundació Mir-Puig, and from Generalitat de Catalunya (AGAUR Grant No.\ 2017 SGR 1341, CERCA program, QuantumCAT \_U16-011424 , co-funded by ERDF Operational Program of Catalonia 2014-2020), MINECO-EU QUANTERA MAQS (funded by State Research Agency (AEI) PCI2019-111828-2 / 10.\allowbreak{}13039/\allowbreak{}501100011033), EU Horizon 2020 FET-OPEN OPTOLogic (Grant No 899794), and the National Science Centre, Poland-Symfonia Grant No.\ 2016/20/W/ST4/00314, Marie Sklodowska-Curie grant STRETCH No.\ 101029393, ``La Caixa'' Junior Leaders fellowships (ID100010434),  and EU Horizon 2020 under Marie Sk\l odowska-Curie grant agreement No 847648 (LCF/BQ/PI19/11690013, LCF/BQ/PI20/11760031,  LCF/BQ/PR20/11770012).
MR and VA acknowledge support from EPSRC/DSTL via MURI award No. EP/N018680/1.
\section*{Appendix. SFA Matrix elements}

In this Appendix, we provide the main steps for calculating the integrals \begin{equation}
\mathcal{I}_1= \int d\mathbf{r}e^{i\mathbf{k}\cdot \mathbf{r}}V(\mathbf{r})
\label{eq:I1}
\end{equation}
and
\begin{equation}
\mathcal{I}_2= \int d\mathbf{r}e^{-i\mathbf{k}\cdot \mathbf{r}}r\cos\theta\psi_{0}(\mathbf{r})
\label{eq:I2}
\end{equation}
in Eqs.~(\ref{eq:SFA_DIR}) and (\ref{eq:SFA2}), for the effective binding potential (\ref{eq:pot}). 
To avoid singularities in the Fourier transform of this potential,
it is multiplied with a damping factor: $V(r)\rightarrow V(r)\times e^{-\epsilon r}$.
Numerical simulations show that this factor does not distort the shape
of the ATI spectrum, although it can shift its overall magnitude; for
this study we have chosen $\epsilon=1.0$. For other methods to treat this singularity see, e.g., Ref.~\cite{Faria2005}.

Using the spatial representation 
\[
\Psi_{0}(\mathbf{r})=R_{n_{0}l_{0}}(r)Y_{l_{0}m_{0}}(\Omega)
\]
of the initial state, where $\Omega$ is the solid angle, Eqs.~(\ref{eq:I1}) and (\ref{eq:I2}) can be simplified as
\begin{eqnarray}
\mathcal{I}_1 &=&-4\pi\left[\frac{1}{\epsilon^{2}+k^{2}}+\frac{a_{1}}{\epsilon^{2}+a_{2}^{2}+k^{2}}+\frac{2a_{3}a_{4}}{(\epsilon^{2}+a_{4}^{2}+k^{2})^{2}}+\frac{a_{5}}{\epsilon^{2}+a_{6}^{2}+k^{2}}\right]\\
\mathcal{I}_2 & =&\frac{(4\pi)^{3/2}}{\sqrt{3}}\sum_{lm}i^{-l}Y_{lm}(\Omega_k)\int drr^{3}R_{n_{0}l_{0}}(r)j_{l}(kr) \int d\Omega Y_{lm}^{*}(\Omega)Y_{10}(\Omega)Y_{l_{0}m_{0}}(\Omega,)\label{eq:fourier_init}
\end{eqnarray}
where we have inserted the damping factor into Eq.~(\ref{eq:pot}), substituted
$\cos\theta$ by a spherical harmonic and used the multipole expansion
for the exponential function in three dimensions. In Eq.~(\ref{eq:fourier_init}), the solid angle $\Omega$ is defined in coordinate space and the solid angle $\Omega_k$ with regard to the intermediate momentum $\mathbf{k}$.

The spatial radial integral
can be solved analytically for an initial state described by a hydrogenic
wave function. For the ground state (1\emph{s}) we use $$R_{10}(r)=2Z^{3/2}e^{-Zr},$$
for the excited state (2\emph{s}) we have $$R_{20}(r)=\frac{Z^{3/2}}{\sqrt{2}}(1-Zr/2)e^{-Zr/2},$$ and
for the polarizable excited state (2\emph{p}) we use $$R_{21}(r)=\frac{Z^{5/2}}{2\sqrt{6}}re^{-Zr/2},$$
where $Z$ is the nuclear charge. 

For the ground state  1\emph{s}, where $l_{0}=0$,
the sum in Eq.~(\ref{eq:fourier_init}) is restricted to $l=1$ and
the spatial integral reads 
as
\[
2Z^{3/2}\int drr^{3}e^{-Zr}\left(\frac{\sin(kr)}{(kr)^{2}}-\frac{\cos(kr)}{kr}\right)=\frac{16Z^{5/2}k}{\left(Z^{2}+k^{2}\right)^{3}}.
\]
Similarly, for 2\emph{s} we obtain 
\[
\frac{Z^{3/2}}{\sqrt{2}}\int drr^{3}e^{-Zr/2}(1-Zr/2)\left(\frac{\sin(kr)}{(kr)^{2}}-\frac{\cos(kr)}{kr}\right)=\frac{2048kZ^{5/2}(2k^{2}-Z^{2})}{\sqrt{2}\left(Z^{2}+4k^{2}\right)^{4}}.
\]

For the excited state 2\emph{p}, where $l_{0}=1$, the sum over $l$
runs from 0 to 2, giving the following three expressions in ascending
order of $l$
\[
\frac{Z^{5/2}}{2\sqrt{6}}\int drr^{3}e^{-Zr/2}\frac{\sin(kr)}{k}=\frac{Z^{5/2}}{2\sqrt{6}}\frac{12Zk(Z/2-k)(Z/2+k)}{k\left(Z^{2}/4+k^{2}\right)^{4}}
\]
\[
\frac{Z^{5/2}}{2\sqrt{6}}\int drr^{3}e^{-Zr/2}\left(\frac{\sin(kr)}{k^{2}r}-\frac{\cos(kr)}{k}\right)=\frac{Z^{5/2}}{2\sqrt{6}}\left(\frac{2(3(Z/2)^{2}-k^{2})}{k\left((Z/2)^{2}+k^{2}\right)^{3}}-\frac{6((Z/2)^{4}-6(Z/2)^{2}k^{2}+k^{4})}{k\left((Z/2)^{2}+k^{2}\right)^{4}}\right)
\]

\[
\frac{Z^{5/2}}{2\sqrt{6}}\int drr^{3}e^{-Zr/2}\left(\frac{3\sin(kr)}{k^{3}r^{2}}-\frac{\sin(kr)}{k}-\frac{3\cos(kr)}{k^{2}r}\right)
=  \frac{Z^{5/2}}{2\sqrt{6}}\left[\mathcal{A}_1(Z,k)-\mathcal{A}_2(Z,k)-\mathcal{A}_3(Z,k)\right],
\]
with $$\mathcal{A}_1(Z,k)=\frac{3Z}{k^{2}\left((Z/2)^{2}+k^{2}\right)^{2}} $$
$$
\mathcal{A}_2(Z,k)= \frac{12Zk(Z/2-k)(Z/2+k)}{k\left((Z/2)^{2}+k^{2}\right)^{4}}
$$
and
$$
\mathcal{A}_3(Z,k)= \frac{3Z((Z/2)^{2}-3k^{2})}{k^{2}\left((Z/2)^{2}+k^{2}\right)^{3}}.
$$

The angular integral of the product of three spherical harmonics
is given by the product of Clebsch-Gordan coefficients ($CG$), whose general form reads as
\[
\int d\Omega Y_{l_{1}m_{1}}^{*}(\Omega)Y_{l_{2}m_{2}}(\Omega)Y_{l_{3}m_{3}}(\Omega)=\sqrt{\frac{(2l_{2}+1)(2l_{3}+1)}{4\pi(2l_{1}+1)}}CG(l_{2},l_{3},l_{1};m_{2},m_{3},m_{1})CG(l_{2},l_{3},l_{1};0,0,0).
\]

If the initial state is perpendicularly polarized, then the angular
part $Y_{l_{0}m_{0}}(\Omega)$  in Eq.~(\ref{eq:fourier_init}) will have some combination of $m_{0}=1,-1$. Therefore, 
the only nonzero contribution to the rescattering term will come
from the element in the summation for which $m=1,-1$. One should note that, so far, the rescattered ATI transition amplitude (\ref{eq:SFA2}) incorporates all possible paths for this three-step process are integrated over, all possible ionization times for the ionization (when the laser is active) and rescattering processes, and all the possible intermediate Volkov states with canonical momentum $\mathbf{k}$. The integrals in  Eq.~(\ref{eq:SFA2}) can be calculated using saddle-point methods, which simplify the numerical effort involved and are easily relatable to an intuitive, orbit-based picture \cite{Salieres2001,Nayak2019,Faria2002}. However, depending on the problem at hand care must be taken.  
If the saddle approximation
is used in all three dimensions, then $k_{x}=0$ and $k_{y}=0$, this constrains the vector
$\mathbf{k}$ to be fixed in the z-direction, leaving no contribution at
all and not enabling meaningful comparisons to the numerical simulation
results to be made. Physically, this implies that only those rescattering processes that occur exactly along the laser polarization axis are included in this approximation, and the remaining processes are selected out. For discussions in the context of molecules see \cite{Chirila2009}. 

We therefore use the saddle point approximation for just two dimensions
in the integral and numerically integrate over the dimension of the
atomic polarization (chosen in the $x$ direction). The action can
be expanded in three dimensions, so that
\[
S(\mathbf{k};t,t')=\frac{1}{2}\intop_{t'}^{t}dt''\left[(k_{z}+A_{z}(t''))^{2}+k_{x}^{2}+k_{y}^{2}\right]
\]
Each variable in the integrand that we approximate is substituted
by its stationary value obtained with the saddle-point approximation, obtained setting the derivative of the action is equal
to zero. This gives
\begin{align}
k_{z}^{st} & =-\frac{1}{t-t'}\intop_{t'}^{t}dt''A_{z}(t'')\label{eq:6}\\
k_{y}^{st} & =0\label{eq:7}.
\end{align}
for $\partial S(\mathbf{k};t,t')/ \partial k_z=0$ and $\partial S(\mathbf{k};t,t')/ \partial k_y=0$, respectively.
For each dimension in which we use the approximation, we make the following substitution
\begin{equation}
\intop d\textbf{k} \left[\frac{2\pi}{\alpha+i(t-t')}\right]^{1/2},\label{eq:8}
\end{equation}
where $\alpha$ is a small parameter to avoid a singularity when $t\rightarrow t'$.
If the saddle point approximation is used in three dimensions, then this
parameter is not necessary unless the initial state has \emph{p} symmetry.
This is because the exponent in the last integral of Eq.~(\ref{eq:SFA2})
goes to zero in the limit when $t\rightarrow t'$ and the spherical
Bessel function of such an argument tends to zero for all orders apart
from for $j_{0}$ - this function only appears in the summation for
an initial \emph{p} state. Thus, modifying (\ref{eq:fourier_init})
with (\ref{eq:6}), (\ref{eq:7}) and (\ref{eq:8}), we obtain 
\begin{align*}
M_\mathrm{Resc}(\mathbf{p}) & =-\intop_{-\infty}^{\infty}dt\intop_{-\infty}^{t}dt'\left[\frac{2\pi}{\alpha+i(t-t')}\right]E(t')e^{iI_{P}t'}e^{-iS(k_{z}^{sp};t,t')}e^{iS(\mathbf{p},t)}\\
& \int dk_{x}e^{-iS(k_{x};t,t')}\times\frac{1}{(2\pi)^{3}}\int d\mathbf{r'}e^{i(\mathbf{k}-\mathbf{p}).\mathbf{r'}}V(\mathbf{r'})\times\frac{1}{(2\pi)^{3/2}}\int d\mathbf{r}e^{-i(\mathbf{k}+\mathbf{A(t')}).\mathbf{r}}\cos\theta\psi_{0}(\mathbf{r})
\end{align*}
where $\mathbf{k}=k_{z}^{sp}\hat{z}+k_{y}^{sp}\hat{y}+k_{x}\hat{x}$.  This specific approach is being used both for parallel and perpendicular bound-state orientation.  This integral was also performed in \cite{Milosevic2013}.

\end{document}